\documentclass[a4paper,english,10pt]{article}
\usepackage[T1]{fontenc}
 \usepackage{ae}
 \usepackage{aecompl}
\usepackage[latin9]{inputenc}
\usepackage{graphicx}
\usepackage{amssymb}
\usepackage{youngtab}
\usepackage{a4wide}

\providecommand{\tabularnewline}{\\}

\newcommand{\SU}[1]{\mathrm{SU}(#1)}
\newcommand{\U}[1]{\mathrm{U}(#1)}
\newcommand{\Sp}[1]{\mathrm{Sp}(#1)}
\newcommand{\eqn}[1]{Eq.(\ref{#1})}
\newcommand{\eqref}[1]{(\ref{#1})}
\usepackage{babel}
\newcommand{\be}{\begin{equation}}
\newcommand{\ee}{\end{equation}}
\newcommand{\ba}{\begin{eqnarray}}
\newcommand{\ea}{\end{eqnarray}}

\newcommand{\Nc}{N}
\newcommand{\Nf}{N_f}

\newcommand{\nq}{M}

\newcommand{\df}{\Delta_f}

\newcommand{\nct}{n}

\newcommand{\lamsun}{\Lambda_{\mbox{\tiny $B$ }}}

\newcommand{\musun}{\mu_{\mbox{\tiny $B$ }}}
\newcommand{\wmg}{W^{\mbox{\scriptsize (mg)}}}
\newcommand{\wel}{W^{\mbox{\scriptsize (el)}}}

\newcommand{\Tr}{\mbox{Tr}}
\newcommand{\asymm}{\tiny\Yvcentermath1\yng(1,1)}
\newcommand{\fund}{\tiny\Yvcentermath1\yng(1)}
\newcommand{\afund}{\tilde{\tiny\Yvcentermath1\yng(1)}}

\begin{document}

\date{\mbox{ }}
\title{{\normalsize  IPPP/09/72 DCPT/09/144\hfill\mbox{}\hfill\mbox{}}\\
\vspace{2.5 cm}
\Large{\textbf{Dual unified $\SU{5}$}}}
\vspace{2.5 cm}
\author{Steven Abel, Valentin V. Khoze\\[3ex]
\small{\em Institute for Particle Physics Phenomenology, Durham University, Durham DH1 3LE, UK}\\[1.5ex]
\small{\em }s.a.abel or valya.khoze@durham.ac.uk\\[1.5ex] }
\date{}
\maketitle

\vspace{2ex}
\begin{abstract}
\noindent Despite the compelling simplicity of minimal supersymmetric unification, the 
proton does not decay as predicted.
A possible explanation is that real unification takes place in a theory which is an electric dual
of the MSSM GUT with messengers, which has
Landau poles below the GUT scale. As was recently pointed out in
Ref.~\cite{AK:Dualification}, unification predictions would be
transmitted across such a duality, but proton decay would be
negligible. In this paper we present electric Seiberg-duals for the
minimal supersymmetric $\SU{5}$ GUT (plus messengers) that
realise this idea, one particularly nice example being an
asymptotically free $\SU{11}\times \Sp{1}^3$ model. We also
discuss how gauge-mediated metastable supersymmetry breaking can be
incorporated into an overall unified picture.

\end{abstract}

\newpage

\maketitle

\section{Introduction}

Landau poles below the GUT scale are a generic feature of theories with direct gauge mediation of supersymmetry (SUSY)
breaking, because the number of messengers is unavoidably large.
We recently proposed in Ref.~\cite{AK:Dualification} that, even in the presence of such Landau poles,
gauge coupling unification can make sense. The less controversial proposal was that the
gauge couplings of the MSSM are "deflected" towards weaker coupling by the messenger sector
(i.e. the SUSY-breaking sector itself in direct mediation)
becoming strongly coupled and entering an electric phase in which there are effectively
fewer messengers. This effect is already evident in direct mediation models \cite{Abel:2007jx}
of the MSSM coupled to a metastable SUSY breaking breaking sector of the
Intriligator-Seiberg-Shih (ISS) type \cite{ISS}.

\begin{figure}[!h]
\begin{center}
\includegraphics[width=80mm]{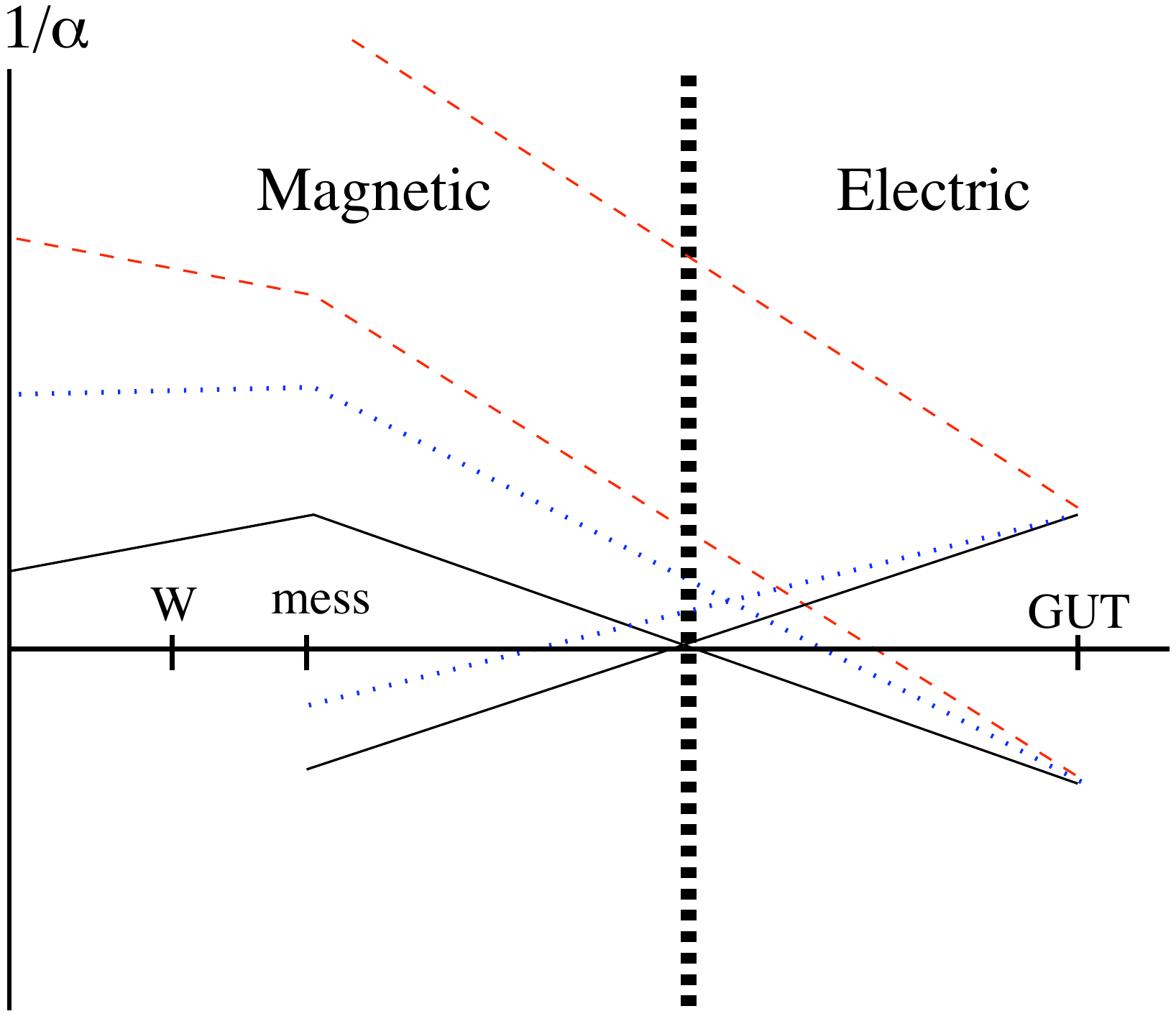}
\caption{\em The dual-unification scenario of Ref.~\cite{AK:Dualification}: the supersymmetric Standard Model appears to run to unphysical gauge unification when there are many messengers in complete $SU(5)$ multiplets. This is mapped to a real unification occurring in an electric dual description that is  valid above the Landau pole scale. \label{fig:dualification}}
\end{center}
\end{figure}

The more controversial proposal, and the one that will be of chief interest in this paper, was that the
Landau poles are real, and that the MSSM is itself a magnetic Seiberg dual \cite{S:Duality,Intriligator:1995au,Strassler:1996ua} of an unknown electric theory.
The argument that unification predictions can still have meaning in this case was based on
the fact that in known examples where a GUT has a dual infra-red (IR)-free magnetic description
and an asymptotically free electric description, the
gauge couplings appear to unify in {\em both} theories at the GUT scale, although in the magnetic theory
this happens at negative $\alpha_{GUT}$ (c.f. Figure \ref{fig:dualification}).
This unphysical gauge unification also happens in the supersymmetric Standard Model with a large number
of messengers in complete $\SU{5}$ representations, and so in \cite{AK:Dualification} it was suggested
that the latter would signify a magnetic dual GUT.

Furthermore it was pointed out that this would explain why the supersymmetric Standard Model appears to unify but the proton does not decay.
This is seen most clearly for proton decay generated by the troublesome "dimension-5" operator in the context of the minimal supersymmetric $\SU{5}$ GUT.
This operator (being generated at the GUT scale) corresponds to
a holomorphic baryon term in the effective superpotential that can only be perturbatively computed
in the electric theory. However the decay takes place at low energies and is therefore computed in the magnetic description with the baryon operator being mapped to the corresponding magnetic baryon. This baryon mapping is well understood and is indeed an important test of the electric/magnetic duality. Consequently the decay rate is suppressed by the many powers of $\Lambda/M_{GUT}$ associated with the mapping of electric to magnetic baryons, where $\Lambda $ is the typical dynamical scale (i.e. the Landau pole scale where the theories are strongly coupled).

The arguments of Ref.~\cite{AK:Dualification} were
based on known examples of Kutasov duality, first introduced and developed in \cite{K:Adj,KS:DKSS,KSS:DKSS,B:2Adj,BS:Theatre,ILS:NewDualities}.
Related studies of these theories were made in Refs.~\cite{Poppitz:1996wp,Poppitz:1996vh,Klein:1998uc,Klein:2003wa,singlets}. Unfortunately none of these models can do a very convincing impression of the conventional $\SU{5}$ GUT, although some rather general
extensions of Kutasov (or more generally Kutasov-Schwimmer-Seiberg (KSS)) duality suggest
that one might be able to get closer to it\footnote{For example Ref.~\cite{singlets}
derived a new class of Kutasov duality in which  gauge and flavour singlets were used to extend the possible particle content to incorporate multiple generations of antisymmetrics, symmetrics and adjoints.}.
(We should add that the idea that there exists an electric dual for the MSSM is an old one, having been  
suggested in Refs.\cite{S:Duality,Klebanov:2000hb}. Perhaps the closest attempts in the
earlier literature are the dual SO(10) models that were presented in Ref.~\cite{Berkooz:1997bb} albeit in a different context.)

In this paper we present duals of the conventional $\SU{5}$ GUT in which the
field content  (in particular the asymmetrics and higgs fields) arise as bound states (mesons) of some product group which also dualize (or confine).
This is a similar procedure
to that of Refs.~\cite{Berkooz:1995km,Pouliot:1995zc,Pouliot:1995me,Pouliot:1995sk} and especially to \cite{Leigh:1997sj},
in which one deconfines the theory into a larger one that can be easily dualized (i.e. it is vector-like with respect to the
Standard Model gauge groups)\footnote{Indeed
the procedure we will follow grew out of discussions with C. Csaki, Y. Shirman and J. Terning
to whom we are extremely grateful.}. This technique has been used to derive dualities between chiral theories and
even between non-chiral and chiral theories.

\begin{figure}[!h]
\begin{center}
\includegraphics[totalheight=0.37\textheight,viewport=0 240 650 750,clip]{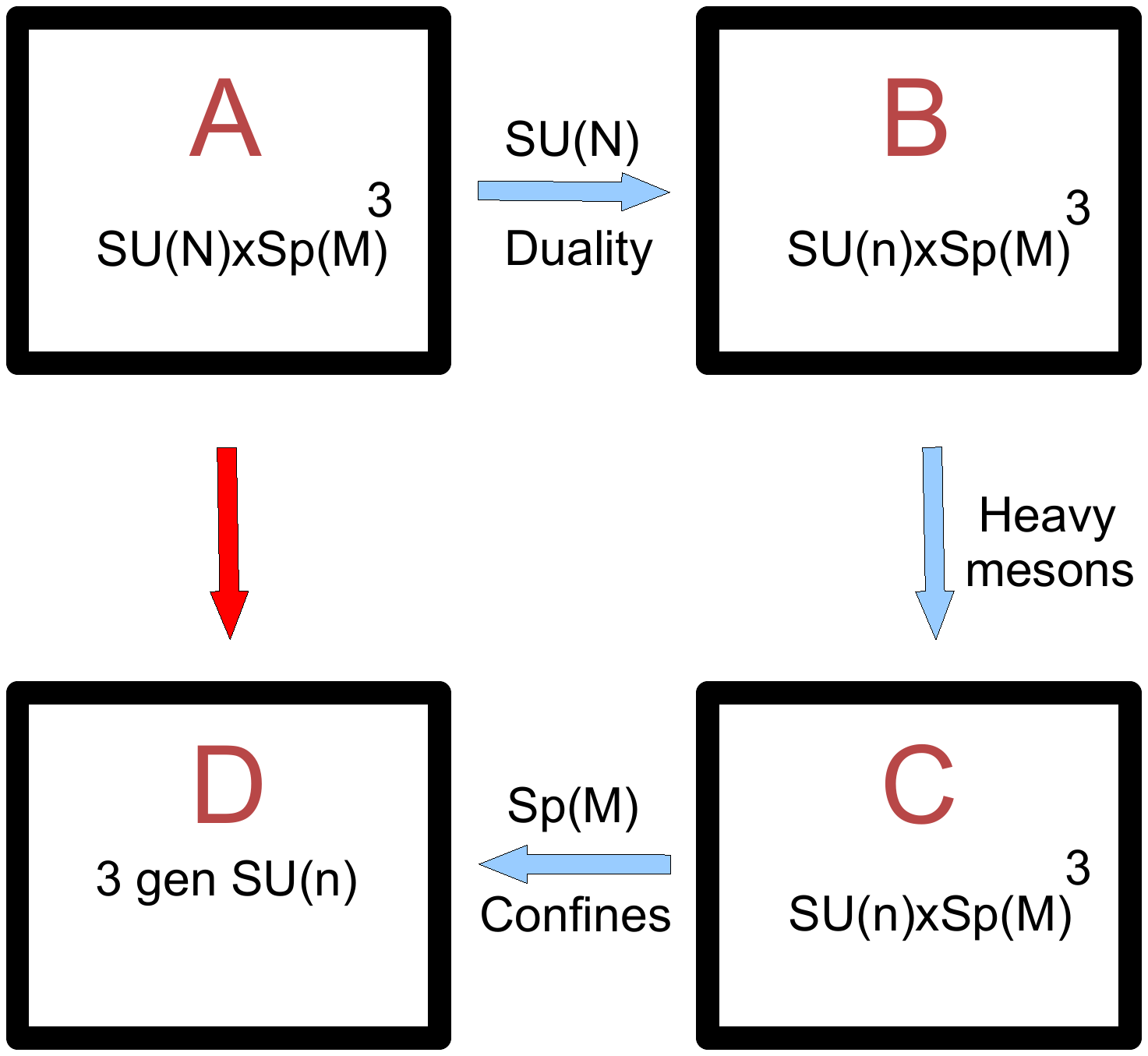}
\caption{\em A route from a supersymmetric $\SU{\Nc}\times \Sp{{M}}^3$ gauge theory, to a magnetic
dual which is the supersymmetric $\SU{5}$ model, in which the Sp groups confine. A similar diagram obtains
for the $\SU{n}$ model with metastable SUSY breaking. The arrows indicate the direction of flow to the IR. Their
labels indicate the type of duality. \label{fig:dualityfig}}
\end{center}
\end{figure}

Before presenting the details, let us summarize how the duality works: the chain of dualities whereby it is established is outlined in
Figure~\ref{fig:dualityfig}.
We begin in the electric-phase (theory A, top left). It has an $\SU{N}$ gauge group coupled to three $\Sp{{M}}$ gauge groups, one for each generation.
This theory has three generations of bifundamental matter fields and
hence a simple vector-like structure\footnote{Chiral theories, such as the minimal $\SU{5}$ GUT are as we have said hard to dualize directly;
typically cancellation of the $\SU{\Nc}^3$ anomaly either restricts
the magnetic gauge group to be also $\SU{\Nc}$ (which is trivial), or requires that any magnetic dual has a different number of quarks,
thereby changing the flavour symmetries and making anomaly matching impossible.}
and of course for the present discussion we also take  ${\cal N}=1$ supersymmetry. In addition one can add pairs of fundamentals and antifundamentals
(which for the sake of argument can be called messengers).  In order to break the GUT symmetries we shall also include an
 adjoint field for the SU groups. For this reason the $\SU{N}$ electric/magnetic dualities will generally be of the Kutasov type, in which a
"dangerously irrelevant" operator is added to the superpotential.
Finally we will (as for the usual $\SU{5}$ model) assume the most general set of leading couplings in the superpotential
consistent with a discrete $R$-parity symmetry in the model. These
couplings will be important in determining the IR behaviour of the theory. In order for this to be a good
electric dual we can choose the parameters (e.g. numbers of messengers) such that both
the gauge factors are asymptotically free.

The $\SU{N}$ gauge group becomes strongly coupled and we can dualize it, to obtain
an $\SU{n}\times \Sp{M}^3$ theory (B), in which $\nct=k \Nf - \Nc$, with $\Nf$ being the effective number of $\SU{\Nc}$ fundamentals.
The quartic matter couplings become
mass terms for the mesons in theory B. Depending on the choice of parameters, some or
all of these will be relevant, and the corresponding mesons may be integrated out yielding theory C, which has the same structure of
couplings as theory A. If enough mesons are lost in this way, the $\Sp{M}$ groups become strongly coupled themselves
and they can in turn be dualized or confine to yield theory D which has different, $\Sp{m}$, gauge groups.
The mesons of this duality are three generations of antisymmetrics and three generations of higgs fields:
this is our putative magnetic dual theory into which we would like to fit the conventional $\SU{5}$ model,
with one pair of higgses playing the usual role, and the two extra generations of higgs pairs remaining unsplit
and heavy (i.e. being essentially just extra pairs of messenger fields).

There are two options for the magnetic theory D.
For a certain choice of parameters the $\Sp{M}$ groups confine (i.e. $m=0$) and
theory D contains {\em only} $\SU{5}$.
This establishes the existence of an electric dual for the minimal supersymmetric $\SU{5}$ GUT with
messenger pairs.  Supersymmetry breaking could then be assumed to occur in a separate sector to which
this sector couples through its messenger fields. On the other hand an interesting possibility is for the $\Sp{M}$
confinement itself to break supersymmetry with no further sectors required. This would be a direct application of the suggestion in Ref.~\cite{ISS}
that S-confining Sp models may break supersymmetry in a metastable vacuum (albeit in a noncalculable way).
We find quite attractive the idea that the same duality is responsible for the appearance of antisymmetrics,
the chiral nature of the conventional $\SU{5}$ model, and for the SUSY  breaking.
A nice feature of the S-confined case is also that
the Yukawa couplings for the up quarks are generated nonperturbatively.
Generally it is rather easy to find examples of this kind, one particularly attractive case being an
asymptotically free $\SU{11}\times\Sp{1}^3$ model dual to the conventional $\SU{5}$ GUT with extra messengers.

A second option is to incorporate
{\em calculable} metastable ISS type SUSY breaking directly
in the $\Sp{M}^3$ half of the gauge group. In this case we let some of
the meson masses in theory B be smaller than the dynamical scale of that theory. These terms then correspond precisely to the
mass-deformations required in the electric phase of the $\Sp{M}$ ISS model \cite{ISS}. The flow from theory C to D
is precisely the ISS Sp duality, and one finds that theory D breaks supersymmetry in a metastable vacuum, with the
$\Sp{m}$ gauge groups now providing the necessary SUSY breaking dynamics. The end result is an electric/magnetic ISS-like
duality between theories A and D, namely a non-chiral  supersymmetric GUT gauge theory and a chiral magnetic
$\SU{n}$ theory directly coupled to (metastable) broken supersymmetry (albeit without the nice feature of a nonperturbatively generated
Yukawa coupling).  Unfortunately this option doesn't work quite so well: it is not possible to build such a calculable SUSY breaking model in this set-up
in which all the gauge groups of theory A are asymptotically free, which seems to indicate that some kind of cascading
behaviour might be difficult to avoid \cite{Klebanov:2000hb}. We will nevertheless present an example of how the metastable SUSY breaking could be
implemented in this case, with the above caveat, and with the possibility that other sorts of GUT (perhaps with antisymmetrics rather than adjoints breaking
the GUT symmetry as in Ref.~\cite{ILS:NewDualities}) may yield better UV behaviour.

There are some interesting phenomenological aspects of these models we should emphasize.
One interesting feature is that the fundamental half of the higgs pair is composite, and there are only antifundamental
higgs fields in the unconfined magnetic theory. Moreover the resulting higgs mass terms (the $\mu$-term of the MSSM so to speak) are related to the confinement scales, and this makes the discussion of higgs masses rather interesting. Moreover any fundamental or anti-fundamental fields that are {\em not} composite under the confining gauge group can only have Yukawa couplings  that are suppressed by the high-energy (e.g. GUT) scale. Non-composite fields would therefore be ideal for playing the role of messengers. The adjoint field can be
chosen to break the $\SU{5}$ in the conventional manner, and yield the MSSM.  Doublet-triplet mass splitting can be implemented
in the usual way and we will discuss its interpretation in the electric theory
as a necessary tuning of VEVs rather than masses, as one would expect (we can't shed any light on this fine-tuning problem however).
As it was anticipated already in Ref.~\cite{AK:Dualification} the model exhibits unification (at one-loop)
and yet the proton decay is completely suppressed.

\section{The dualities in detail}

We now reconstruct the chain of dualities described in the introduction and depicted in figure \ref{fig:dualityfig}.
Note that the theory we outline below is not the only choice of parameters. One can in fact choose any number of
$\Sp{M}$ factors, extra $\Sp{M}$ fundamentals in the electric theory and so on. Our main aim here is to establish
that $\SU{5}$ GUTs can be fairly simply dualized so for the moment we will usually make the most minimal choices.

\subsection{\em Theory A}
It is in fact easier to begin with theory A, shown in table~\ref{deconf1}\footnote{The dualities we will be using are well established and
satisfy all the usual tests of for example `t Hooft anomaly matching. Therefore we will not burden the reader by including all
the global charges.}.
The model is based on an ${\cal N}=1$ supersymmetric $\SU{N}\times \Sp{M}^3$ gauge theory (one $\Sp{M}$ for each generation),
with an adjoint field for the $\SU{N}$ gauge group denoted $X$.
We assume three generations of bifundamentals $Y_a$ coupling the $\SU{N}$ group to each of the $\Sp{M}$ groups.
To have an anomaly free vector-like theory we have to introduce $6M$ antifundamentals of
$\SU{N}$.
In addition we may add an arbitrary number of massless $\Sp{M}$ fundamentals (which are singlets under $\SU{N}$): for concreteness
and simplicity of presentation we add one field, $Z_{a}$, for each $\Sp{M}_{a=1\ldots 3}$. Later we will allow
for more general numbers of Sp fundamentals.
\begin{table}[htdp]
\caption{\em Theory A: the electric $\SU{\Nc}\times \Sp{M}^3$ model. We also allow  $\Delta_f$ pairs of
fundamental/antifundamental (under $\SU{\Nc}$).}
\begin{center}
\begin{tabular}{|c||c|c|c|c|c|}
\hline
$ $ & $\SU{\Nc}$ & $\Sp{M}_a$  & $R_p$ \tabularnewline
\hline
\hline
$Y_{a=1\ldots 3}$ & $\fund$ & $\fund$ & $i$  \tabularnewline
\hline
$ \tilde{Q}_{\bar{J}=1\ldots 3\nq}$ & $\afund$  & 1 & $1$  \tabularnewline
\hline
$ \tilde{H}_{\bar{J}=1\ldots 3M}$ & $\afund$  & 1 & $-1$  \tabularnewline
\hline
$ \tilde{F}_{\bar{J}=1\ldots \Delta_f }$ & $\afund$  & 1 & $-i$  \tabularnewline
\hline
$ {F}_{{J}=1\ldots \Delta_f}$ & $\fund$  & 1 & $i$  \tabularnewline
\hline
$X$ & Adj &  1 & 1 \tabularnewline
\hline
\hline
$Z_a$ & 1& $\fund $ & $i$ \tabularnewline
\hline
\end{tabular}
\par\end{center}
\label{deconf1}
\end{table}%

One should bear in mind that for eventual phenomenology one would like to have a discrete symmetry
such as $R$-parity in the model to distinguish between matter fields and higgses. We assign
$R$-parity $i$ to the bifundamentals $Y_a$, $R$-parity $+1$  to $3 \nq$ of the fundamentals which
we call matter fields, $\tilde{Q}_{\bar{J}=1\ldots 3 \nq}$, and
$R$-parity $-1$ to the rest of the fundamentals which we call higgses $\tilde{H}$. (These charges
are the opposite from the usual MSSM ones, but in the eventual magnetic theory they will be he right way round.)
We also add $\Delta_f$ of $\SU{N}$-fundamental/antifundamental pairs (extra messenger fields)  $F$ and $\tilde{F}$
with $R$-parity $\pm i$ respectively.

KSS duality requires a non-zero superpotential for the $\SU{N}$ adjoints which we take to be cubic.
Again, other values are
possible but the higher you go the more difficult it becomes to achieve duality (in the sense that the gauge groups
quickly become unwieldy).
$\SU{N}$ GUT symmetry breaking will be driven by lower order deformations to the $X^3$ superpotential which we can take to be a
mass term for $X$. We write down the
most general leading (modulo powers of $X$ which will get large VEVs) terms
for the remaining fields consistent with $R$-parity and the gauge symmetries:
\ba
W_A &=& \frac{m_X}{2} X^2+\frac{s_0}{3}X^3 + \kappa_{i} Z  YX^i  \tilde{H} \nonumber \\
&& \hspace{-0.4cm} +
\lambda_{ij} \tilde{Q}X^i Y Y X^j \tilde{H} +
\lambda'_{ij} \tilde{F}X^i Y Y X^j \tilde{F}
+
 \lambda''_{ij} \tilde{H}X^iF  \tilde{Q}X^jF
 + \lambda'''_{ij}\tilde{F}X^iF  \tilde{F}X^jF
 \, . \label{WAone}
\ea
where the couplings carry dimensions. Later we discuss the dimensions in more detail, but for the 
moment we simply assume that the typical couplings would be
$\kappa_{i}\sim   M_{P}^{-i}$ and
$\lambda_{ij}\sim M_{P}^{-(i+j+1)}$ where $M_P$ is some fundamental scale not too far above the 
GUT scale. We have suppressed the generation indices on the couplings $\kappa_{ia\bar{J}}$,
$\lambda_{ija\bar{I} \bar{J}}$,
$\lambda^{\prime}_{ij a  \bar{I}\bar{J}}$,
$\lambda^{\prime\prime}_{ij \bar{J}J  \bar{I}I}$ and
$\lambda^{\prime\prime\prime}_{ij \bar{J}J  \bar{I}I}$  -- there is no reason for them to be diagonal.
The truncation of the chiral ring (see below) means that we need only include
terms upto $X$  in the superpotential so $i,j=0,1$ only. (More precisely one could include higher order
terms in $X$ in the electric theory, but these can be identified with the lower ones in the magnetic theory through the 
$X$ equations of motion \cite{KSS:DKSS}.)

\subsection{\em Theory B}

One can straightforwardly determine when theory A has a magnetic dual description.
We will briefly recapitulate the required duality which is a particular ($k=2$)
version of the duality developed by Kutasov, Schwimmer and Seiberg (KSS) \cite{K:Adj,KS:DKSS,KSS:DKSS}.
The electric KSS theory in their general notation consists of an $\SU{N}$ gauge theory
with $\Nf$ flavours of fundamental/antifundamental pairs, an adjoint $X$, and a superpotential that
takes the form
\be
W_{\mbox{\tiny KSS}}^{({\mbox{\tiny elec}})}=\sum_{i=0}^{k-1}\frac{s_{i}}{k+1-i}X^{k+1-i}+{\lambda} X,
\label{Wksselectr}
\ee
where $k$ is an integer parameter. The deformations (i.e. the $s_{i>0}$
terms) are responsible for breaking the GUT symmetry so it is natural to take 
$s_i \sim M_{GUT}^{2+i-k}$. (Note that the parameter ${\lambda}$ is a Lagrange multiplier
to fix $\Tr(X)=0$.) Indeed the $F_{X}$-term
equation for non-zero $s_{i}$'s can easily be solved by diagonalizing
the $X$ using $\SU{\Nc}$ rotations; the equation for a single entry
$X$ on the diagonal is
\be
W'=0\equiv\sum_{i=0}^{k-1}s_{i}X^{k-i}+{\lambda}\, .
\ee
This is a $k$'th order polynomial so there are $k$ roots: hence
\be
\langle X\rangle=\left(\begin{array}{ccc}
a_{1}\mathbf{I}_{r_{1}}\\
 & a_{2}\mathbf{I}_{r_{2}}\\
 &  & ...a_{k}\mathbf{I}_{r_{k}}\end{array}\right)\,\,\,\,;\,\,\,\sum_{i=1}^{k} r^{k}=\Nc
 \ee
When there is only the $X^{k+1}$ term the $\SU{N}$ symmetry obtains, but adding the
lower $i>0$ terms (the deformations) breaks the gauge symmetry as \be
\SU{\Nc}\rightarrow \SU{r_{1}}\times \SU{r_{2}}\ldots \SU{r_{k}}\times \U{1}^{k-1}.\ee
These equations of motion truncate the chiral ring, so that $X^k$ can be related to
lower order operators, allowing a matching to be made to a magnetic dual theory.

The unbroken magnetic gauge group is $\SU{\nct}$ where $\nct=k \Nf-\Nc$.
An equivalent form of superpotential and breaking pattern
holds for the magnetic theory.
To determine the magnetic theory it is convenient to define a shifted field
$X_s=X+b \mathbf{1} $ to ensure that the $X^k$ term in the superpotential vanishes.
The superpotential \eqref{Wksselectr} becomes
\be
W_{\mbox{\tiny KSS}}^{({\mbox{\tiny elec}})}=\sum_{i=0}^{k-1}\frac{t_{i}}{k+1-i}X_s^{k+1-i}+{\lambda} X_s
\ee
where the coefficient $t_{1}=0$ vanishes. We can then identify a set of composite fields playing
the role of mesons. There are $k$ of them;
\begin{equation}
\Phi_{j}=\tilde{Q}X_{s}^{j}Q\,\,;\,\,\, j=0\ldots k-1\, ,
\end{equation}
where $Q$ and $\tilde{Q}$ stand generically for all the $\SU{N}$
fundamentals and antifundamentals, i.e. $Y,F$ and $\tilde{Q},\tilde{H},\tilde{F}$ in table~\ref{deconf1}.
We will generally adopt
the convention that magnetic fields and parameters are denoted by small letters while those of the
electric theory are denoted by capitals.
The superpotential in the magnetic theory is, with this convention,
\begin{equation}
W_{\mbox{\tiny KSS}}^{({\mbox{\tiny mag}})}=\sum_{i=0}^{k-1}\frac{-{t}_{i}}{k+1-i}x_{s}^{k+1-i}+\frac{1}{\musun^{2}}\sum_{l=0}^{k-1}t_{l}\sum_{j=1}^{k-l}\Phi_{j-1}\tilde{q}x_{s}^{k-j-l}q\label{eq:wmag}\, ,\end{equation}
where $q$ and $\tilde{q}$ stand for the corresponding $\SU{n}$
fundamentals and antifundamentals. The parameter ${\musun}$ is
expected to be of order the dynamical scale of $\SU{N}$ in theory $B$, $\lamsun$ and encodes our ignorance about
the matching between the dynamical scales in the electric and magnetic theories.

In Ref.~\cite{KSS:DKSS} it was shown that the vacuum structure of the two theories matches, and indeed this
form of superpotential can be fixed by considering the $R$-symmetries of the model and performing
anomaly matching with the couplings included as states in the spectrum \cite{singlets}.
In the broken theory the $\SU{r_i}$ subfactors
are generically mapped to each other as per usual SQCD duality: i.e. $\SU{r_i}\leftrightarrow\SU{\bar{r}_i}$ where
$\bar{r}_i=\Nf-r_i$. For the purposes of this paper we shall be mostly interested
in the minimal $\SU{5}$ models which break $\SU{5}\rightarrow \SU{3}\times \SU{2}\times \U{1}$
and so will focus on $k=2$.

In the above, $\mu_B$
is the parameter governing the matching between the electric and
magnetic KSS theories (i.e. theories A and B respectively) as
follows \cite{KSS:DKSS}: \be \Lambda_A^{b_{\tiny{\SU{\Nc}}}^{(A)}}
\Lambda_B^{b_{\tiny{\SU{\nct}}}^{(B)}} = \left(
\frac{\mu_B}{t_0}\right)^{b_{\tiny{\SU{\Nc}}}^{(A)}+
b_{\tiny{\SU{\nct}}}^{(B)}}\, . \ee Note that the parameter
$M_{GUT}$ then appears in the matching of the subtheories if the
GUT symmetry is broken. That is, if the breaking is \ba
\SU{\Nc} &\rightarrow &  \SU{r_{1}}\times \SU{r_{2}}\ldots \SU{r_{k}}\times \U{1}^{k-1}\nonumber \\
\SU{\nct} &\rightarrow &  \SU{\bar{r}_{1}}\times
\SU{\bar{r}_{2}}\ldots \SU{\bar{r}_{k}}\times \U{1}^{k-1}\, , \ea
then the SQCD subtheories are matched with the electric
and magnetic dynamical scales for the $i$'th factor being related
as \be \label{sooper} \Lambda_{A,i}^{b^{(A)}_i}
\Lambda_{B,i}^{b^{(B)}_i}=\mu_i^{b^{(A)}_i+b^{(B)}_i} \ee where
\be 
\label{droppy}
\mu_i \sim \frac {\mu_B^2 }{t_0 M_{GUT}}\, \equiv \bar{\mu}. \ee
Eq.\eqref{sooper} and the fact that the $\mu_i $ are all
approximately degenerate (depending only on the masses of GUT
states that are integrated out) is enough to ensure dual
unification, i.e. unification occuring in both theories, as
discussed in Ref.~\cite{AK:Dualification}.

Before continuing we should for later use also 
mention that one can choose to normalize the mesons in the obvious way. 
By comparison with standard SQCD duality, the relation in \eqref{droppy} 
is equivalent to the normalized mesons (denoted by a bar)
\begin{equation}
\label{normalization}
\bar{\Phi}_{j}=\frac{t_0 M_{GUT}^{1-j}}{\mu_B^2}\tilde{Q}X_{s}^{j}Q\,\,;\,\,\, j=0\ldots k-1\, ,
\end{equation}
so that for example we get the usual identification $\bar{\mu} \bar{\Phi}_{0} = \tilde{Q}{Q}$. 
The superpotential in the magnetic theory is then 
\begin{equation}
W_{\mbox{\tiny KSS}}^{({\mbox{\tiny mag}})}=M_{GUT}^3\sum_{i=0}^{k-1}\frac{-\bar{t}_{i}}{k+1-i}\bar{x}_{s}^{k+1-i}+\sum_{l=0}^{k-1}\bar{t}_{l}\sum_{j=1}^{k-l}\bar{\Phi}_{j-1}\tilde{q}\bar{x}_{s}^{k-j-l}q\label{eq:wmag2}\, ,\end{equation}
where bars over $\bar{t}_l$ and $\bar{x}_s$ indicate that they are now in $M_{GUT}$ units and dimensionless. We will use this normalization
procedure if we need to discuss for example the physical masses of the mesons, but we shall not show it explicitly. 

Now, in terms of the
theory with general $k$ and $N_f$ effective flavours the RG behaviour is as follows: the $b_0$-coefficients
of the beta-functions for the electric and magnetic theories are given by
\ba
b_{\mbox{\tiny $\SU{\Nc}$}}   &=&   2 \Nc - N_f \, \nonumber \\
{\bar{b}}_{\mbox{\tiny $\SU{\Nc}$}}  & = &  (2k-1) \Nf - 2\Nc  \, .
 \ea
Hence in the range
\be
\label{constraint40}
\frac{2k\Nc}{2k-1} > kN_f > \Nc+1
\ee
the electric theory is asymptotically free and the magnetic theory is an IR free
dual description. (Equality, $kN_f = N+1$, corresponds to S-confinement). In the range
\be
2 N > N_f > \frac{2N}{2k-1}
\ee
the (unbroken) theory flows to an IR fixed point and has two equivalent
descriptions. Note that in the broken theory (i.e. when $\SU{N}\rightarrow \SU{r_1}\times \ldots \SU{r_k}\times \U{1}^{k-1}$)
the separate $\SU{r_i}$ factors may be in different regimes, but the vacuum stability bound $kN_f > N+1 $ remains. Thus
the range
\be
\label{konst1}
 kN_f > \Nc+1\, ,
\ee
defines where a dual theory exists. This is the criterion we have to use. (In particular
note that at the next link in the chain of dualities we will create more fundamentals of $\SU{n}$
so it would be meaningless to use the IR free criterion.)
In the specific $k=2$ case under consideration we have
$\Nf=6M+\df$ and this corresponds to
\be
\label{constraint1}
 6M+\df> \frac{\Nc+1}{2}
 \, .
 \ee
 In this range theory A
is either asymptotically free or runs to a fixed point, and in either case there is an equivalent dual description, theory B.
The content of theory B is as shown in table \ref{modelb},
where
\be
n= k (6 M +\df) - \Nc\, .
\ee
\begin{table}[htdp]
\caption{\em Theory B: the intermediate $\SU{\nct}\times \Sp{M}^3$ model with $\nct=12 M +2\df - \Nc\,$. Note that the
flavour identification of electric/magnetic quarks is $\tilde{Q}\leftrightarrow \tilde{h}$ and $\tilde{H}\leftrightarrow \tilde{q}$.}
\begin{center}
\begin{tabular}{|c||c|c|c|c|c|}
\hline
$ $ & $\SU{\nct}$ & $\Sp{M}_a$  & $R_p$ \tabularnewline
\hline
\hline
$y_a$ & $\fund$ & $\fund$ & $-i$  \tabularnewline
\hline
$\tilde{h}_{\bar{J}=1\ldots 3 M}$ & $\afund$  & 1 & $1$  \tabularnewline
\hline
$ \tilde{q}_{\bar{J}=1\ldots 3 M}$ & $\afund$  & 1 & $-1$  \tabularnewline
\hline
$ \tilde{f}_{\bar{J}=1\ldots \Delta_f}$ & $\afund$  & 1 & $i$  \tabularnewline
\hline
$ {f}_{{J}=1\ldots \Delta_f}$ & $\fund$  & 1 & $-i$  \tabularnewline
\hline
$x$ & Adj &  1 & 1 \tabularnewline
\hline
\hline
$Z_a$\, $\Phi_{i\,aJ}$  & 1 & $\fund$  & $i$ \tabularnewline
\hline
 $\chi_{i\,a\bar{J}}$  & 1 & $\fund$  & $-i$ \tabularnewline
\hline
 $\Sigma_{i\,a\bar{J}}$  & 1 & $\fund$  & $1$ \tabularnewline
\hline
$(\phi_H)_i\equiv(FX^i \tilde{H})$   & 1 &1   & $-i$ \tabularnewline
\hline
$(\phi_Q)_i \equiv(F X^i \tilde{Q})$   & 1 &1   & $i$ \tabularnewline
\hline
$(\phi_F)_i \equiv(F X^i\tilde{F})$   & 1 &1   & $1$ \tabularnewline
\hline
\end{tabular}
\par\end{center}
\label{modelb}
\end{table}%
The $6M(6M+\df) $ mesons, $ (\Phi_{i=0,1})_{a\bar{J}}= Y_a X^i \tilde{Q}_{\bar{J}}\, $, $ (\Sigma_{i=0,1})_{a\bar{J}}= Y_a X^i \tilde{F}_{\bar{J}}\, $
and $ (\chi_{i=0,1})_{a\bar{J}}= Y_a X^i \tilde{H}_{\bar{J}}\, $
are $2(6M+\df) $ fundamentals for each $\Sp{M}$ and radically change the running of those
gauge couplings. What happens next depends on the superpotential which among other things
determines whether these fields are integrated out or not.

With the above rules, the required shift is $X_s=X+\frac{m}{2s_0}\mathbf{1}$ (i.e. $b=\frac{m}{2s_0}$).
The magnetic superpotential is (noting that we identify the quark flavours as $\tilde{Q}\leftrightarrow \tilde{h}$
and $\tilde{H}\leftrightarrow \tilde{q}$)
then found to be
\ba
W_B &=&
 \frac{{m}_x}{2} x^2-\frac{s_0}{3}x^3 + \kappa_{i} Z  \chi_i  \nonumber \\
&& \hspace{-0.4cm} +
\breve{\lambda}_{ij} \Phi_i \chi_j
+
\breve{ \lambda}'_{ij} \Sigma_{i} \Sigma_{ j}+
\breve{ \lambda}''_{ij} \phi_{H\,i} \phi_{Q\, j} + \breve{ \lambda}'''_{ij} \phi_{F\,i} \phi_{F\, j}  \nonumber \\
&&  \hspace{-0.4cm}  + \frac{s_0}{\musun^2}
\left(
\Phi_i \tilde{h} x_s^{1-i} y +
\chi_i \tilde{q} x_s^{1-i} y +
\Sigma_i \tilde{f} x_s^{1-i} y \right. \nonumber \\
&& \left.
\hspace{0.4cm}+\hspace{0.2cm} \phi_{Q\,i} \tilde{h} x_s^{1-i} f +
\phi_{H\,i} \tilde{q} x_s^{1-i} f +
\phi_{F\,i} \tilde{f} x_s^{1-i} f
\right)
\, ,
\ea
where repeated generation indices are summed,  and where we have 
defined the shifted parameters
\ba
x_s & =& x+\breve{b}\mathbf{1} \nonumber \\
\breve{b} & = & -\frac{{m}_x}{2} \nonumber \\
{m}_x &=& \frac{\Nc}{\nct} m_X\nonumber \\
\breve{\lambda}_{00} &=& \lambda_{00} - b \lambda_{01} - b \lambda_{10}+ b^2 \lambda_{11} \nonumber \\
\breve{\lambda}_{01} &=& \lambda_{01} -  b \lambda_{11}  \nonumber \\
\breve{\lambda}_{10} &=& \lambda_{10} - b \lambda_{11}  \nonumber \\
\breve{\lambda}_{11} &=& \lambda_{11}\, ,
\ea
and similar for $\breve{\lambda}'$, $\breve{\lambda}''$ and  $\breve{\lambda}'''$. Again we have suppressed all except the $i,j$ indices
and we use the un-normalized convention for the mesons.

\subsection{\em Theory C}

The $\kappa $, $\breve{\lambda}$, $\breve{\lambda}'$, $\breve{\lambda}''$,  $\breve{\lambda}'''$ terms are masses for the mesons. Using the 
physical normalization in \eqref{normalization} and assuming $t_0\sim 1$, we find their masses to be typically of order
\ba
m_{\rm meson} &\sim &  \lambda_{ij} \mu_B^4 M_{GUT}^{i+j-2} \nonumber \\
&\sim &  \frac{\mu_B^4}{M_{GUT}^3} \left( \frac{M_{GUT}}{M_P}\right)^{i+j+1} \nonumber \\
&\sim &  \frac{\bar{\mu}^2}{M_{GUT}} \left( \frac{M_{GUT}}{M_P}\right)^{i+j+1}
\, ,
\ea
where we recall from \eqn{droppy} that $\bar{\mu}=\mu_B^2/t_0M_{GUT}$.
Heavy mesons can be integrated out, but to be consistent we should check that their mass terms are relevant operators in the IR,
(which is not automatically the case).
In order to check this we can use the non-anomalous exact $R$-charges of the mesons
at the IR fixed point of the undeformed theory (i.e. the theory without Yukawa couplings) \cite{S:Duality}. These are 
\be
R_{\Phi_i }= 2-\frac{4}{k+1}\frac{\Nc}{\Nf}+ \frac{2i}{k+1} \, ,
\ee
and similar for $\chi_i$.
For all the mass terms to be relevant we need only check that the
highest dimension term is less than 3: this term is of the form $\Phi_1 \chi_1$
and its dimension is $2\times \frac{3}{2} R_{\Phi_1\chi_1}$. Thus
all mass terms are relevant if $N_f<\frac{4}{5}\Nc$ or
\be
\label{constraint4}
M <\frac{4\Nc-5\df}{30} \, .
\ee
Assuming that the rank is maximal, all the mesons except a single linear combination
of $Z_a$ and  $\Phi_{i\, a\bar{J}}$ for each $\Sp{M}$  get masses. We will refer to these
massless fundamentals of $\Sp{M}$ as $z_a$.
Once the heavy mesons are integrated out the spectrum reduces to that in table~\ref{modelcp}
and the superpotential reproduces the equivalent
couplings to those that were present in theory A:
\ba
W_C&=& \frac{m_x}{2} x^2-\frac{s_0}{3}x^3 + \tilde{\kappa}_{i}\, z  (yx_s^i  \tilde{h})
\nonumber \\
&& +\hspace{0.1cm}
\tilde{\lambda}_{ij}\, (\tilde{q}x_s^i y)( y x_s^j \tilde{h})
+\tilde{ \lambda}'_{ij}\, ( \tilde{f}x_s^i y)( y x_s^j \tilde{f})
+ \tilde{\lambda}''_{ij} \, (\tilde{h}x_s^if)(  \tilde{q}x_s^jf) + \tilde{ \lambda}'''_{ij}\,(\tilde{f}x_s^if)(  \tilde{f}x_s^jf)
 \, ,
\ea
where we bracket $\SU{n}$ singlets. For completeness
we can present the couplings in a slightly simplified case where
$\kappa_{i}$ is small enough for the remaining light meson $z$ to be approximated as almost pure $Z$:
\ba
\tilde{\lambda}_{ij} & \approx & -\frac{s_0^2}{\mu_B^4}(\breve{\lambda}^{-1})_{1-j, 1-i} \,\, ; \,\, \tilde{\lambda}'_{ij}  = - \frac{1}{4} \frac{s_0^2}{\mu_B^4} (\breve{\lambda}^{\prime\prime-1})_{1-i, 1-j}
\,\, ; \,\,
\tilde{\kappa}_i \approx - \frac{s_0}{\mu_B^2} (\breve{\lambda}^{-1})_{1-i,j}  \kappa_j
 \nonumber \\
\tilde{\lambda}''_{ij} & = & - \frac{s_0^2}{\mu_B^4}(\breve{\lambda}^{\prime\prime-1})_{1-j, 1-i} \,\, ; \,\, \tilde{\lambda}'''_{ij}  = - \frac{1}{4}\frac{s_0^2}{\mu_B^4} (\breve{\lambda}^{\prime\prime\prime -1})_{1-i, 1-j}  \,\, ,
\ea
where the inverse refers to the $2\times 2$ matrices corresponding to the $i,j$ indices only. In the normalized 
basis \eqref{normalization}, the $x_s$ would appear with bars (i.e. in dimensionless units of $M_{GUT}$) and the couplings 
would then be typically of order the inverse meson mass,
\be 
\bar{\tilde{\lambda}}\sim m_{\rm meson}^{-1}\, .
\ee
\begin{table}[htdp]
\caption{\em Theory C: the low energy intermediate $\SU{\nct}\times \Sp{M}^3$ model with $\nct=12 M +2\df - \Nc\,$.}
\begin{center}
\begin{tabular}{|c||c|c|c|c|c|}
\hline
$ $ & $\SU{\nct}$ & $\Sp{M}_a$  & $R_p$ \tabularnewline
\hline
\hline
$y_a$ & $\fund$ & $\fund$ & $-i$  \tabularnewline
\hline
$\tilde{h}_{\bar{J}=1\ldots 3 M}$ & $\afund$  & 1 & $1$  \tabularnewline
\hline
$ \tilde{q}_{\bar{J}=1\ldots 3 M}$ & $\afund$  & 1 & $-1$  \tabularnewline
\hline
$ \tilde{f}_{\bar{J}=1\ldots \Delta_f}$ & $\afund$  & 1 & $i$  \tabularnewline
\hline
$ {f}_{{J}=1\ldots \Delta_f}$ & $\fund$  & 1 & $-i$  \tabularnewline
\hline
$x$ & Adj &  1 & 1 \tabularnewline
\hline
\hline
$z_a$  & 1 & $\fund$  & $i$ \tabularnewline
\hline
\end{tabular}
\par\end{center}
\label{modelcp}
\end{table}%

This reproduction of couplings is the main point to appreciate. In particular it is easy to see that
it occurs for any choice of discrete symmetries. That is, the $\lambda$-couplings in theory A in \eqref{WAone}
are determined by
the symmetries;  the $\SU{N}$ invariants then turn into mesons in theory B, for which the quartic couplings become mass terms;
when the mesons are integrated out in theory C, each mass term reproduces the corresponding bilinear operator of composite magnetic mesons.
Conversely if a coupling is absent in theory A, then that mesons never get massive and the corresponding
coupling is also absent from theory C.

\subsection{\em Theory D and D$'$}


Now, we would like to be in the regime where the $\Sp{M}_a$ groups
can either have an IR-free magnetic dual description, or can be S-confined to get the final theory. We call these
two possibilities D and D$'$ respectively.

Generally, for the $\Sp{M}_a$ gauge factors in Theory C to have a dual description in
the free magnetic window\footnote{That is, the $\Sp{m}_a$ of the Theory D are IR-free.}, one requires
\cite{Intriligator:1995ne}
\be
\label{constraint2a} {\frac{3}{2} (M+1)} >
{N_{f\,{\Sp{M}}}} \geq M+2
 \, ,
 \label{freemagw}
\ee
where we used the standard notation that the effective number of $\Sp{M}$ flavours,
$N_{f\,{\Sp{M}}}$, is half the
actual number of $\Sp{M}$ fundamentals $\fund$ i.e.
\be
N_{f\,{\Sp{M}}} := \, \frac{1}{2}\,N_{\fund_{\Sp{M}}} \ ,
\qquad  N_{\fund_{\Sp{M}}}=\, n+1 \, .
\ee
Here $n$ represents the number of $y_a$ flavours of each $\Sp{M}_a$ gauge group, and
the 1 corresponds to the single massless fundamental $z_a$, see table~\ref{modelcp}.
Equation \eqref{freemagw} then gives
\be
\label{constraint2c}
{3 (M+1)} > n+1   \geq  2M+4
 \, .
\ee
In this window there exists an IR free
magnetic dual $\Sp{m}$ description with
\ba
m &=&  N_{f\,{\Sp{M}}} - (M+2) \nonumber \\
&=& \frac{n-3}{2}-M   \nonumber \\
& = & 5M  + \df - \frac{N+3}{2}
\ea
whose mesons include antisymmetrics of $\SU{n}$.
The spectrum is shown in table~\ref{modeld},
\begin{table}[htdp]
\caption{\em Theory D: the generic low energy $\SU{\nct}\times \Sp{m}^3$ model with ${M}=\frac{n-3}{2}-m  $.}
\begin{center}
\begin{tabular}{|c||c|c|c|c|c|}
\hline
$ $ & $\SU{\nct}$ & $\Sp{m}_a$  & $R_p$ \tabularnewline
\hline
\hline
$\tilde{y}_a$ & $\afund$ & $\fund$ & $i$  \tabularnewline
\hline
$\tilde{h}_{J=1\ldots 3M}$ & $\afund$  & 1 & $1$  \tabularnewline
\hline
$ \tilde{q}_{\bar{J}=1\ldots 3M}$ & $\afund$  & 1 & $-1$  \tabularnewline
\hline
$ \tilde{f}_{\bar{J}=1\ldots \Delta_f}$ & $\afund$  & 1 & $i$  \tabularnewline
\hline
$ {f}_{{J}=1\ldots \Delta_f}$ & $\fund$  & 1 & $-i$  \tabularnewline
\hline
$x$ & Adj &  1 & 1 \tabularnewline
\hline
\hline
$\zeta_a$ & 1 &  $\fund $ & $i$ \tabularnewline
\hline
\hline
$a_a$ & $\asymm$ & 1 & $-1$ \tabularnewline
\hline
$h_a$ & $\fund$ & 1& $1$ \tabularnewline
\hline
\end{tabular}
\par\end{center}
\label{modeld}
\end{table}%
where the mesons, $a_a \equiv y_a y_a $  and $h_a \equiv z_a y_a $, give us the necessary
representations of antisymmetrics for the minimal $\SU{n}$ GUT. Note that the bifundamentals $\tilde{y}_a$
in the dual theory have their flavour charges (i.e. their $\SU{n}$ charges) reversed and that the $R$-parities of the
matter and higgs fields are the conventional ones for a $\SU{5}$ GUT-like model.

Equality on the {\it r.h.s} of \eqref{freemagw} (i.e. $ N_{f\,{\Sp{M}}} = M+2$ ) corresponds to S-confinement of $\Sp{M}$. This is the
regime where the dual magnetic $\Sp{m}_a$ gauge groups are trivial, $m=0$. The spectrum of this S-confined theory
is shown in table~\ref{modeldp}.
\begin{table}[htdp]
\caption{\em Theory D$'$: the spectrum of the confined low energy $\SU{\nct}$ model when $m=0$ or  $M = \frac{n-3}{2}  $.}
\begin{center}
\begin{tabular}{|c||c|c|c|c|}
\hline
$ $ & $\SU{\nct}$   & $R_p$ \tabularnewline
\hline
\hline
$\tilde{h}_{J=1\ldots 3M}$ & $\afund$   & $1$  \tabularnewline
\hline
$ \tilde{q}_{\bar{J}=1\ldots 3M}$ & $\afund$   & $-1$  \tabularnewline
\hline
$ \tilde{f}_{\bar{J}=1\ldots \Delta_f}$ & $\afund$   & $i$  \tabularnewline
\hline
$ {f}_{{J}=1\ldots \Delta_f}$ & $\fund$   & $-i$  \tabularnewline
\hline
$x$ & Adj  & 1 \tabularnewline
\hline
\hline
$a_a$ & $\asymm$  & $-1$ \tabularnewline
\hline
$h_a$ & $\fund$ & $1$ \tabularnewline
\hline
\end{tabular}
\par\end{center}
\label{modeldp}
\end{table}%

One can check that anomalies cancel as they should: the contribution to $\SU{n}^3$ anomalies is
$
-6(M+m) + 3(n-4) +3 =0\, .
$
The superpotential is derived from $W_{C}$ with the required additional meson terms. For generic values of $m$ it takes
the form
\be
W_D = \frac{{m}_x}{2} x^2-\frac{s_0}{3}x^3+ \tilde{\kappa}_{i} \, h x_s^i  \tilde{h}+
\tilde{\lambda}_{ij} \, \tilde{h} x_s^i a x_s^j \tilde{q} +
\tilde{\lambda}'_{ij} \, \tilde{f} x_s^i a x_s^j \tilde{f} +
\frac{1}{\mu_D}
(a \tilde{y} \tilde{y}+ h \zeta \tilde{y})+\mbox{quartic} + W_{D}^{\rm dyn}\, ,
\label{supE}
\ee
while for the S-confining theory ($m=0$) it is
\be
W_{D'} = \frac{{m}_x}{2} x^2-\frac{s_0}{3}x^3+ \tilde{\kappa}_{i}\, h x_s^i  \tilde{h}+
\tilde{\lambda}_{ij} \, \tilde{h} x_s^i a x_s^j \tilde{q} +
\tilde{\lambda}'_{ij} \, \tilde{f} x_s^i a x_s^j \tilde{f} +\mbox{quartic} + W_{D'}^{\rm dyn}\, ,
\label{supEprime}
\ee
These expressions for the superpotentials of the IR-free and the S-confined theories include
dynamically generated non-perturbative
contributions $W_{D^\prime}^{\rm dyn}$.
In the following subsection we will show that the dynamical superpotential for the S-confined theory
with GUT group $\SU{5}$ is nothing but the Yukawa interaction required for the up-type quark masses,
\be
\SU{5} \, : \qquad W_{D'}^{\rm dyn}=\, \hat{\lambda}\,a a h \, =: W_{\rm up-Yukawa}
\label{up-Yuk}
\ee
The other Yukawa coupling necessary for giving masses to down-type quarks is already present
in the superpotentials \eqref{supE},\eqref{supEprime} and is given by $\tilde{\lambda}_{ij}$ with $i=0=j$;
\be
\label{d-Yuk}
W_{d-Yukawa} = \lambda_{d\, ij} \tilde{h} \bar{x}^i_s a \bar{x}^i_s \tilde{q} \, 
\ee 
where 
\be 
\lambda_{d \, ij}  \sim \Lambda_C \bar{\tilde{\lambda}}_{ij} 
\, \sim \, \frac{\Lambda_C}{m_{\rm meson}} \, .
\ee
Note that $m_{\rm meson}$ are the masses of the heavy mesons that we have integrated out:
this ensures that $\lambda_d \lesssim 1$. Corrections to the leading couplings are related to the GUT symmetry breaking, coming 
from the higher adjoint contributions in $\bar{x}_s$ and can induce flavour changing Yukawas.
On the other hand, as we shall see presently, the up-type Yukawas can have no adjoint contribution and so a defining feature
of their generation by nonperturbative effects is that they are diagonal.
Of course both of these Yukawa interactions are required by Standard Model phenomenology in any viable $\SU{5}$ GUT,
and we find it remarkable that the up-type Yukawa interactions which are apparently missing from the `perturbative' superpotential
are generated nonperturbatively by the S-confining Sp gauge dynamics\footnote{If this was not the case one would have
to induce this interaction via an appropriate higher-dimensional operator in the electric dual theory resulting in negligibly small
Yukawa couplings $\hat{\lambda}$, suppressed by inverse powers of the new physics scale, presumably near to $M_{P}$ and, by assumption, above
$M_{GUT}$.}.

We also note that the $\fund$-components of the low energy higgs fields, namely $h_a$, are actually composite, and that
since we are thinking of theory D' as the minimal $\SU{5}$ GUT, then the
$\tilde{\kappa}$ is also a required phenomenological  coupling, namely the
so called higgs $\mu_{\mbox{\tiny MSSM}}$-term. Again using the physical normalization \eqref{normalization} 
we find
\be 
\mu_{\mbox{\tiny MSSM}, \, i} \sim \, \bar{\mu} \left( \frac{\Lambda_C}{m_{\mbox{\tiny meson} }}\right)_{1-i,j}  \left(\frac{M_{GUT}}{M_P}\right)^j \, .
\ee
There are of course three of them, one for each generation of higgs -- in this theory the lowest would then be tuned to split the 
higgs doublets and triplets in the usual manner (of which more later), while the two remaining higgs generations would get large masses.

\subsection{Dynamical Superpotential and up-Yukawa interaction in Theory D'}

Integrating out the $\Sp{M}_a$ degrees of freedom in Theory C generates
non-perturbative superpotentials in the magnetic duals D and D'. We will show that it is related to Yukawa couplings in the
S-confining theories and for this reason we will concentrate here on theory D'.

The exact dynamically generated superpotential can be derived from symmetry considerations following the classic approach of Affleck, Dine and
Seiberg~\cite{Affleck:1983rr,Affleck:1983mk,Affleck:1984xz,Intriligator:1995au}, with the overall coefficient fixed by an instanton computation. For the case at hand where the $\Sp{M}$ gauge dynamics in
Theory C is S-confined in Theory D', the superpotential is given by \cite{ISS}
\be
W_{D'}^{\rm dyn}=\, -\frac{{\rm Pf}({\rm mesons})}{\Lambda_{\rm Sp}^{n-3}}\, ,
\label{Wdyn2}
\ee
where $\Lambda_{\rm Sp}$ is the dynamical scale of the $\Sp{M}$ theory (for out particular case this is $\Lambda_C$), 
$n$ is the number of colours of $\SU{n}$ and the Pfaffian
is over the matrix of mesons emerging from integrating out the $\Sp{M}$ degrees of freedom. As already discussed in the previous section,
these mesons are $(y y):= \Lambda_{\rm Sp} a $  and $(z y) :=\Lambda_{\rm Sp} h$, cf table \ref{modeldp}, where the brackets
denote the symplectic contraction so that mesons are Sp-singlets, and where we have reinstated the dynamical scale $\Lambda_{\rm Sp}$
in the normalization
of the fields. To visualise this expression we first consider a determinant of the
$(n+1)\times(n+1)$ meson matrix,
\be
{\rm det}_{(n+1)\times(n+1)}({\rm mesons}) = \varepsilon^{(n)} \,(zy)^2\,(yy)^{n-1}\, \varepsilon^{(n)}\, ,
\ee
where $\varepsilon^{(n)}$ denote epsilon-symbols contracting $y$ fields to form an $\SU{n}$-singlet. Clearly each elementary
field must appear twice in the determinant, hence there is a factor of $z^2$, and the number of $y$-fields is fixed by the overall
dimension of the determinant $2(n+1)$. Since the Pfaffian is essentially the square root of the determinant above,
we have
\be
W_{D'}^{\rm dyn}=\, -\frac{\varepsilon^{(n)} \,(zy)\,(yy)^{\frac{n-1}{2}}}{\Lambda_{\rm Sp}^{n-3}}
=\, -\frac{1}{\Lambda_{\rm Sp}^{\frac{n-5}{2}}}\, \varepsilon^{(n)} \,h \,a^{\frac{n-1}{2}}\, .
\label{Wdyn3}
\ee
For the $\SU{5}$ GUT the dependence on $\Lambda_{\rm Sp}$ disappears and the superpotential is
cubic in fields $\sim a a h$ giving the up-type Yukawa coupling in Eq.~\eqref{up-Yuk} with
$\hat{\lambda}$ generically of order $1$ (it of course depends on the precise normalisation of canonical fields
in the Kahler potential).

If the number of colours of the $\SU{n}$ theory is greater than 5, the non-perturbative superpotential is
a higher-dimensional operator. However, assuming that $\SU{n}$ can be higgsed down to $\SU{5}$ by giving VEVs to some
of the antisymmetric flavours $a$, the Yukawa structure of the resulting $\SU{5}$ group can be recovered.
For example one can think of $\SU{7} \longrightarrow \SU{5}$ by giving a VEV to one of the $a$ fields;
then $ W_{D'\, \SU{7}\to \SU{5}}^{\rm dyn} \sim \frac{\langle a \rangle}{\Lambda_{\rm Sp}}\,h\,a\,a$.


\subsection{On the nature of the GUT}

Depending on the assumptions we make
and in particular the parameters we choose, the eventual GUT  may
be precisely theory A, or it may deviate from it in a number of ways.
Before presenting explicitly some cases of interest, let us first enumerate
the different types of GUT that may eventually emerge;
some examples will be presented in the following section:

\begin{enumerate}

\item{  The GUT is theory A:}

If the $b_0$ coefficients of the beta functions of theory A are all positive (or zero) then the
theory is asymptotically free and will serve as the weakly-coupled (in the UV) GUT.

\item{Heavy mesons and renormalizable couplings:}

The minimal deviation is that theory A is a low energy approximation to an even larger
theory that has more $Z$-like fields. In other words at higher energies one could
imagine  "integrating in" some heavy mesons just as in theory B. Again assuming maximal rank for all the couplings in theory A, the full spectrum
 is then given by table~\ref{modela}, where the new mesons are identified by their hats. This model is similar in
 structure to theory B, but with different gauge groups. However note that we could choose to "integrate
 in" a different number of mesons, if for example some of the masses were order $M_{GUT}$. The
 superpotential is of the form
  \begin{table}[htdp]
\caption{\em Theory A$_1$: a high energy $\SU{\Nc}\times \Sp{M}^3$ model when heavy mesons are integrated in.}
\begin{center}
\begin{tabular}{|c||c|c|c|c|c|}
\hline
$ $ & $\SU{\Nc}$ & $\Sp{M}_a$  & $R_p$ \tabularnewline
\hline
\hline
$Y_a$ & $\fund$ & $\fund$ & $-i$  \tabularnewline
\hline
$\tilde{H}_{\bar{J}=1\ldots 3 M}$ & $\afund$  & 1 & $1$  \tabularnewline
\hline
$ \tilde{Q}_{\bar{J}=1\ldots 3 M}$ & $\afund$  & 1 & $-1$  \tabularnewline
\hline
$ \tilde{F}_{\bar{J}=1\ldots \Delta_f}$ & $\afund$  & 1 & $i$  \tabularnewline
\hline
$ {F}_{{J}=1\ldots \Delta_f}$ & $\fund$  & 1 & $-i$  \tabularnewline
\hline
$X$ & Adj &  1 & 1 \tabularnewline
\hline
\hline
$\hat{Z}_a$\, $\hat{\Phi}_{i\,aJ}$  & 1 & $\fund$  & $i$ \tabularnewline
\hline
 $\hat{\chi}_{i\,a\bar{J}}$  & 1 & $\fund$  & $-i$ \tabularnewline
\hline
 $\hat{\Sigma}_{i\,a\bar{J}}$  & 1 & $\fund$  & $1$ \tabularnewline
\hline
$(\hat{\phi}_H)_i\equiv(FX^i \tilde{H})$   & 1 &1   & $-i$ \tabularnewline
\hline
$(\hat{\phi}_Q)_i \equiv(F X^i \tilde{Q})$   & 1 &1   & $i$ \tabularnewline
\hline
$(\hat{\phi}_F)_i \equiv(F X^i\tilde{F})$   & 1 &1   & $1$ \tabularnewline
\hline
\end{tabular}
\par\end{center}
\label{modela}
\end{table}%
\ba
W_{A_1} &=&
 \frac{{m}_X}{2} X^2+\frac{s_0}{3}X^3 +\hat{ \kappa}_{i} Z \hat{ \chi}_i  \nonumber \\
&& \hspace{-0.4cm} +
\hat{\lambda}_{ij}{\hat \Phi}_i\hat{ \chi}_j
+
\hat{ \lambda}'_{ij} \hat{\Sigma}_{i} \hat{\Sigma}_{ j}+
\hat{ \lambda}''_{ij} \hat{\phi}_{H\,i} \hat{\phi}_{Q\, j} + \hat{ \lambda}'''_{ij} \hat{\phi}_{F\,i} \hat{\phi}_{F\, j}  \nonumber \\
&&  \hspace{-0.4cm}  + \frac{s_0}{\mu_A^2}
\left(
\hat{\Phi}_i \tilde{H} X^{1-i} Y +
\hat{\chi}_i \tilde{Q} X^{1-i} Y +
\hat{\Sigma}_i \tilde{F} X^{1-i} Y \right. \nonumber \\
&& \left.
\hspace{0.4cm}+\hspace{0.2cm} \hat{\phi}_{Q\,i} \tilde{H} X^{1-i} F +
\hat{\phi}_{H\,i} \tilde{Q} X^{1-i} F +
\hat{\phi}_{F\,i} \tilde{F} X^{1-i} F
\right)
\, .
\ea

\item{A chiral GUT-like electric dual:}

The next possibility is that the $\Sp{M}$'s of theory $A$ or $A_1$ become strongly coupled and one dualizes to
a new electric theory $A_2$ with antisymmetrics, that closely mirrors the $\SU{5}$ magnetic theory.
It is easy, by referring to the C-B duality, to write down this alternative electric theory, A$_2$, which is appropriate if
 theory A (or the obvious extension to A$_1$) has
 \be
 \label{constraint3}
 N_{f\,{\Sp{M}}} = \frac{N+1}{2} \geq  (M+2)\, .\ee
We then find a dual theory with gauge group $\SU{N}\times \Sp{\hat{M}}^3$ where
\ba
\hat{M} &=&  N_{f\,{\Sp{M}}}  - (M+2) \nonumber \\
&=& \frac{N-3}{2}-M \, .
\ea
The spectrum (in table~\ref{modela2}) and superpotential are trivially read off from theory C; we find
\ba
W_{A_2} & = & \frac{{m}_X}{2} X^2+\frac{s_0}{3}X^3+ {\kappa}_{i} \, H X^i  \tilde{H} \nonumber \\
&& +
{\lambda}_{ij} \, \tilde{H} X^i A X^j \tilde{Q} +
{\lambda}'_{ij} \, \tilde{F} X^i A X^j \tilde{F} +
\frac{1}{\mu_A}
(A \hat{Y} \hat{Y}+ H \hat{Z} \hat{Y})+\mbox{quartic}\, ,
\ea
where $A_a=Y_aY_a$ and $H_a=Z_a Y_a $. As in the previous theory the superpotential is
renormalizable apart from the terms with higher powers of $X$.
\begin{table}[htdp]
\caption{\em Theory A$_2$: the generic high energy $\SU{\Nc}\times \Sp{\hat{M}}^3$ model when the
$\Sp{M}$'s become strongly coupled again in the UV, with ${M}=\frac{N-3}{2}-\hat{M}  $.}
\begin{center}
\begin{tabular}{|c||c|c|c|c|c|}
\hline
$ $ & $\SU{\Nc}$ & $\Sp{\hat{M}}_a$  & $R_p$ \tabularnewline
\hline
\hline
$\hat{Y}_a$ & $\afund$ & $\fund$ & $-i$  \tabularnewline
\hline
$\tilde{Q}_{J=1\ldots 3M}$ & $\afund$  & 1 & $1$  \tabularnewline
\hline
$ \tilde{H}_{\bar{J}=1\ldots 3M}$ & $\afund$  & 1 & $-1$  \tabularnewline
\hline
$ \tilde{F}_{\bar{J}=1\ldots \Delta_f}$ & $\afund$  & 1 & $i$  \tabularnewline
\hline
$ {F}_{{J}=1\ldots \Delta_f}$ & $\fund$  & 1 & $-i$  \tabularnewline
\hline
$X$ & Adj &  1 & 1 \tabularnewline
\hline
\hline
$\hat{Z}_a$ & 1 &  $\fund $ & $-i$ \tabularnewline
\hline
\hline
$A_a$ & $\asymm$ & 1 & $-1$ \tabularnewline
\hline
$H_a$ & $\fund$ & 1& $-1$ \tabularnewline
\hline
\end{tabular}
\par\end{center}
\label{modela2}
\end{table}%
This is a different possible dual for the minimal $\SU{5}$ model of theory D (or the confined theory without $\Sp{m}$ groups, D$'$)
the direct relation between the parameters being given by
\ba
m &=& 2\Nc+\df -5\hat{M}-9 \nonumber \\
\nct &=& 5\Nc+2\df -12\hat{M}-18\, .
\ea

\item{The GUT is actually theory C (deflected unification):}

One can find (many) examples in which all the beta functions of theory C are already
both negative so that it runs to weak coupling in the UV. Assuming that the Sp groups are responsible for SUSY breaking
then this is an example of deflected unification. In other words a change in the dynamics of the
hidden sector removes the apparent Landau poles in the visible sector by reducing the
effective number of visible flavours in the UV.

\item{An infinite cascade of dualities:}

Finally, an unwelcome option that we would like to avoid is that the $\SU{N}$ group of theory A$_2$ hits another Landau pole
below the GUT scale: this would happen if $\hat{M}$ were large, which means that
the $\SU{N}$ group has effectively acquired many extra flavours driving its beta
function positive again.
This chain of dualities is reminiscent of a cascade and we suspect that is what would occur,
although it is not possible to go further because we are unable to dualize the
(chiral) theory A$_2$ very simply (i.e. it is no easier than the initial problem of finding
the dual of the $\SU{5}$ GUT).  Note however that $\hat{M}$ is not generally large and so
this cascading behaviour is by no means generic and can be easily avoided.

\end{enumerate}

\subsection{Remarks on matching}

To close this section we should remark on the utility and necessity of the deconfinement technique. This will
also allow us to introduce the idea of baryon matching which will be of importance later when discussing proton decay.

First we emphasize that the matching of each pair of theories in the chain from A to D satisfies all the usual tests of
matching of moduli spaces and global anomalies. This is straightforward to achieve
between adjacent theories in the chain, for example theories B and C, but  it would not be
possible between D and A directly. To see why consider the matching of the
baryons and antibaryons of theory B and theory C when $\df=0$: the electric antibaryons are defined
schematically as
\begin{eqnarray}
\tilde{B} & = & \tilde{Q}^{\Nc}\end{eqnarray}
with the obvious contraction of colour indices, and similar for magnetic
baryons. In order to match the degrees of freedom of the moduli-space we define dressed quarks as \begin{eqnarray}
\tilde{Q}_{l} & = & X^{l}\tilde{Q}\\
\tilde{H}_{l} & = & X^{l}\tilde{H}\,\,\,\,\,\,\,\,\,\,\,\,\,\,\,\,\,\,\,\mbox{$l=1\ldots k-1$},\end{eqnarray}
and similar for magnetic quarks, where we can consider arbitrary $k$.
The matching is explicitly as follows. The electric antibaryon is
\be
\tilde{B}^{(n_{0}\ldots m_{k-1})}=(\tilde{Q}_{0})^{n_{0}}\ldots(\tilde{Q}_{k-1})^{n_{k-1}}(\tilde{H}_{0})^{m_{0}}\ldots(\tilde{H}_{k-1})^{m_{k-1}}
\ee
where \be
\sum_{i=0}^{k-1}n_{i}+m_{i}=\Nc.
\ee
There are \begin{eqnarray}
\sum_{\{\mathbf{n},\mathbf{m}\}}\left(\begin{array}{c}
3\nq \\
n_{0}\end{array}\right)\ldots\left(\begin{array}{c}
3\nq\\
n_{k-1}\end{array}\right)\left(\begin{array}{c}
3M \\
m_{0}\end{array}\right)\ldots\left(\begin{array}{c}
3M \\
m_{k-1}\end{array}\right) & = & \left(\begin{array}{c}
6 Mk \\
\Nc\end{array}\right)\\
\\\end{eqnarray}
of them. In the magnetic antibaryon, the antifundamentals are
mapped to their magnetic counterparts as $\bar{n}_{i}=3M-n_{k-i-1}$
and $\bar{m}_{i}=3M-m_{k-i-1}$, where
\be
{\tilde{b}}^{(\bar{n}_{0}\ldots\bar{m}_{k-1})}=(\tilde{h}_{0})^{\bar{n}_{0}}\ldots(\tilde{h}_{k-1})^{\bar{n}_{k-1}}(\tilde{q}_{0})^{\bar{m}_{0}}\ldots(\tilde{q}_{k-1})^{\bar{m}_{k-1}}.
\ee
Note that \begin{eqnarray}
\sum_{i=0}^{k-1}\bar{n}_{i}+\bar{m}_{i} & = & 6kM-\left(\sum_{i=0}^{k-1}n_{k-i-1}+m_{k-i-1}\right)\\
 & = & 6kM -\Nc=\nct\end{eqnarray}
as required for the contraction. The number of magnetic baryon degrees of freedom is
\begin{eqnarray}
\sum_{\{\mathbf{n},\mathbf{m}\}}\left(\begin{array}{c}
M\\
\bar{n}_{0}\end{array}\right)\ldots\left(\begin{array}{c}
M\\
\bar{n}_{k-1}\end{array}\right)\left(\begin{array}{c}
M\\
\bar{m}_{0}\end{array}\right)\ldots\left(\begin{array}{c}
M\\
\bar{m}_{k-1}\end{array}\right) & = & \left(\begin{array}{c}
6kM\\
\nct\end{array}\right)\\
 & = & \left(\begin{array}{c}
6kM \\
\Nc\end{array}\right)\end{eqnarray}
precisely matching the electric ones. Moreover $\tilde{B}^{(n_{0}\ldots m_{k-1})}$
transforms in totally antisymmetric representations of $\SU{M}_{\tilde{Q}}$
and $\SU{M}_{\tilde{H}}$, so that the precise correspondence is
\be
\tilde{b}^{(\bar{n}_{0}\ldots\bar{m}_{k-1})}\leftrightarrow\varepsilon^{(3M)}\ldots\varepsilon^{(3M)}\tilde{B}^{(n_{0}\ldots m_{k-1})}
\ee
with the Levi-Cevita contractions leaving the required flavour
representations in the magnetic baryon (since for example $\varepsilon^{(3M)}$ acting
on an $n_{i}$ index totally antisymmetric representation of $\SU{M}_{\tilde{Q}}$
gives a $3M-n_{i}=\bar{n}_{k-i-1}$ dimensional representation).
(Incidentally, it is this identification that forces the negation of the flavour
charges between electric and magnetic fundamentals.) Furthermore all
the $\U{1}$ charges match.

Even for theories A and D this matching would work for the antibaryons.
However the baryon matching would be much more
difficult. In the intermediate theories B and C the matching for baryons would go as per the antibaryons
above but would involve $y$'s. Moreover on confinement (or dualizing) these turn
into either $a$'s or $h$'s. Hence
baryon matching between theory D and some electric dual theory A would require quarks
dressed with $a$'s as well as adjoints: but one would have to consider unlimited strings
of antisymmetrics and adjoints because the chiral ring is not truncated and this makes the direct
matching impossible to achieve.

\section{Example models}

Having derived the general form of the expected duality, one can search for examples that exhibit
the different types of behaviour. Here we present some of the
interesting examples, but will not undertake an exhaustive survey
(as we noted already, all types of behaviour described above can be found, but not necessarily for
$\SU{5}$ as will become clear). We will allow for arbitrary number of light SU singlets transforming in the
fundamental of Sp in both theory A and theory C, defined to be $N_{\rm Sp}$ and $n_{\rm Sp}$ for each $\Sp{M}$ group respectively
(so that the case above with a single $Z_a$ and $z_a$ corresponds to $n_{\rm Sp}=N_{\rm Sp}=\frac{1}{2}$).
Increasing $N_{\rm Sp}$ corresponds to integrating in Sp fundamentals in the manner described in the
previous section, whereas increasing $n_{\rm Sp}$ corresponds to  not integrating out the (Sp fundamental) mesons
whose mass falls below the scale $\Lambda_C$ of theory C. Note that in theory $D$ there are $6n_{\rm Sp}$ fundamental
higgses created as mesons of the Sp duality.

For convenience we first summarize the A$\leftrightarrow$D duality and all of the
relevant constraints in terms of the parameters of theory A (namely $\df$, $N$, ${M}$, $N_{\rm Sp}$ and $n_{\rm Sp}$).
The duality between theory A with gauge group $\SU{N}\times \Sp{{M}}^3$ and
D with $\SU{n}\times \Sp{m}^3$ is
\ba
m &=& 5M+n_{\rm Sp}+\df-2-\frac{N}{2}  \nonumber \\
\nct &=& 12M+2\df-N\, .
\ea
The beta functions in the two theories are
\ba
b^{\tiny (A)}_{\mbox{\tiny $\SU{\Nc}$}}   &=&   2 \Nc - 6M- \df \, \nonumber \\
{b}^{\tiny (D)}_{\mbox{\tiny $\SU{\nct}$}}  & = & 9-6n_{\rm Sp}- \nct - \df \, \nonumber \\
&= &  \Nc-12 M +9-6n_{\rm Sp}- 3\df \nonumber \\
b^{\tiny (A)}_{\mbox{\tiny $\Sp{{M}}$}}   &=&   3{M+1}-\frac{(\Nc+2N_{\rm Sp})}{2}  \, \nonumber \\
{b}^{\tiny (D)}_{\mbox{\tiny $\Sp{m}$}}  & = & 3(m+1)-\frac{(\nct+2n_{\rm Sp})}{2}  \, \nonumber \\
&= & 9 {M} -\Nc +2\df+2n_{\rm Sp}-3
 \, .\ea
Then the constraints we apply to for example get an asymptotically free
theory A and an IR-free theory D are
\ba
b^{\tiny (A)}_{\mbox{\tiny $\SU{\Nc}$}} & > & 0 \nonumber \\
b^{\tiny (A)}_{\mbox{\tiny $\Sp{{M}}$}} & > & 0 \nonumber \\
b^{\tiny (D)}_{\mbox{\tiny $\SU{\nct}$}} & < & 0 \nonumber \\
b^{\tiny (D)}_{\mbox{\tiny $\Sp{m}$}} & < & 0 \, ,
\ea
as well as the previous constraints which were in descending order:
\ba
N+2N_{\rm Sp}  & \geq & 2M+4  \nonumber \\
6M+\df  &>& \frac{N+1}{2} \nonumber \\
M &<& \frac{4N-5\df}{30 }\nonumber \\
{3 (M+1)}  > n+2n_{\rm Sp}   & \geq & 2M+4  \, .
\ea
The first of these is relevant if one wishes to dualize to a theory A$_2$
as described above. The last of these (the IR-freedom of $\Sp{m}$) of course ensures the
$b^{\tiny (D)}_{\mbox{\tiny $\Sp{m}$}}  <  0$ condition.
Equality in the first and last indicates the case where both theory A and theory C S-confine.
We now present some examples of the different types of GUT outlined above ("S-confined" below
always refers to the $\Sp{m}$ groups).\\

\begin{itemize}

\item{\it \underline{ S-Confined $\SU{5}$ model dual to $\SU{11}\times \Sp{1}^3$ GUT}: [with $\df=2$, $n_{\rm Sp}=\frac{1}{2}$, $N_{\rm Sp}=\frac{1}{2}$]   }
\vspace{0.2cm}

This is an attractive 2 messenger pair + 3 higgs pair model and is the optimal case which has an asymptotically free
electric phase ($b^{\tiny (A)}_{\mbox{\tiny $\SU{\Nc}$}}=14$, $b^{\tiny (A)}_{\mbox{\tiny $\Sp{{M}}$}}=0$) and an IR free
magnetic phase ($b^{\tiny (D)}_{\mbox{\tiny $\SU{\Nc}$}}=-1$). (One can also take $N_{\rm Sp}=0$ in which case $b^{\tiny (A)}_{\mbox{\tiny $\Sp{{M}}$}}=\frac{1}{2}$.)
Note that the beta functions of the broken magnetic theory
are always more positive than the unbroken theory (i.e. the Landau poles are at lower energy than in the unbroken theory).
Interestingly this case is {\em also} an example of a deflected unification theory since the beta functions of
theory C are $b^{\tiny (C)}_{\mbox{\tiny $\SU{\Nc}$}}=2$ and $b^{\tiny (C)}_{\mbox{\tiny $\Sp{{M}}$}}=3$. The GUT could be either theory.

The reader may be wondering why the beta functions of theory C are different from those of
theory D since S-confinement is not supposed to change them. The reason is that the confinement leads to
new fundamental higgses in the theory that are paired up with the antifundamental $\tilde{h}$'s in mass terms
(i.e. the $\kappa $-terms in the superpotential). Below their masses one can integrate out the
3 higgs pairs from theory D, and the beta functions indeed then return to those of theory C as they should. Thus very
generally we observe that in the confining case the deflection is always in the right direction (i.e. the direction of less
effective flavours) since the change in beta functions on going from theory D to C must be the same as the change
in beta functions on integrating out the higgses from theory D.

\begin{table}[htdp]
\caption{\em The IR theory corresponding to a $\SU{11}\times \Sp{1}^3$ GUT with two generations of messengers.}
\begin{center}
\begin{tabular}{|c||c|c|c|c|}
\hline
$ $ & $\SU{5}$   & $R_p$ \tabularnewline
\hline
\hline
$\tilde{h}_{J=1\ldots 3}$ & $\afund$   & $1$  \tabularnewline
\hline
$ \tilde{q}_{\bar{J}=1\ldots 3}$ & $\afund$   & $-1$  \tabularnewline
\hline
$ \tilde{f}_{\bar{J}=1\ldots 2}$ & $\afund$   & $i$  \tabularnewline
\hline
$ {f}_{{J}=1\ldots 2}$ & $\fund$   & $-i$  \tabularnewline
\hline
$x$ & Adj  & 1 \tabularnewline
\hline
\hline
$a_a$ & $\asymm$  & $-1$ \tabularnewline
\hline
$h_a$ & $\fund$ & $1$ \tabularnewline
\hline
\end{tabular}
\par\end{center}
\label{modeldperf}
\end{table}%

\item{\it \underline{S-Confined $\SU{5}$ model dual to $\SU{15}\times \Sp{1}^3$ GUT}: [with $ \df=4$, $n_{\rm Sp}=\frac{1}{2}$, $N_{\rm Sp}=\frac{1}{2}$]   }
\vspace{0.2cm}

This is a confined 4 messenger pair + 3 higgs pair model. The UV model is asymptotically free only for
the $\SU{15}$ with the beta functions being
$b^{\tiny (A)}_{\mbox{\tiny $\SU{\Nc}$}}=20$, $b^{\tiny (A)}_{\mbox{\tiny $\Sp{{M}}$}}=-2$
and $b^{\tiny (D)}_{\mbox{\tiny $\SU{\nct}$}}=-3$. The $\Sp{1}^3$ groups therefore
grow in the UV, but of course this can be acceptable if the associated Landau pole is chosen to
be above $M_{GUT}$.

\item{\it \underline{S-Confined $\SU{5}$ deflected to unconfined $\SU{5}\times \Sp{1}^3$}: [with $ \df=4$, $n_{\rm Sp}=\frac{1}{2}$, $N_{\rm Sp}=\frac{1}{2}$]   }
\vspace{0.2cm}

This model has an S-confined Sp group with a minimal $\SU{5}$ spectrum as in table \ref{modeldperf}, with
4 messenger pairs and $b^{\tiny (D)}_{\mbox{\tiny $\SU{\nct}$}}=-3$. (Note that together with the extra higgs pairs
theory D has effectively 6 messenger pairs.) The unconfined model has
$b^{\tiny (C)}_{\mbox{\tiny $\SU{\nct}$}}=0$ and $b^{\tiny (A)}_{\mbox{\tiny $\Sp{{M}}$}}=3$ so this is an example of deflection.
Note that the difference of $3$ in the beta function is because of the composite higgs states which are induced in the magnetic theory D as mesons of the
Sp duality. As per our comment above, once these are integrated out below their masses the
beta functions of theory D will return to those of theory C. Nevertheless, the main point here is that the beta function
increases in the UV direction in going from theory C to D.

\item{\it \underline{ IR-free $\SU{9}\times\Sp{1}^3$ dual to
$\SU{43}\times \Sp{4}^3$ GUT}: [with $ \df=2$, $n_{\rm Sp}=\frac{5}{2}$, $N_{\rm Sp}=\frac{1}{2}$]  } \vspace{0.2cm}

In this model there are $2 \times \df =4$ $\Sigma$-mesons in theory B (for each Sp factor),
we keep all of them light and as such they remain in theory C
after {\it heavy} mesons have been integrated out of theory B.
Since the $\Sigma$-mesons are charged under $\Sp{M}$ they play the role of additional fundamental
Sp flavours and take part in the Sp duality to theory D. It then follows that
the resulting theory D contains $\Sp{1}$ factors which are IR-free and can serve as the weakly coupled SUSY-breaking sector.
In other words, the Sp sectors which were already required to deconfine antisymmetric flavours and to enable the
Seiberg duality of SU factors, in this example provide for the ISS type metastable supersymmetry breaking \cite{ISS}.
Supersymmetry breaking due to Sp factors in the IR will be examined in more
detail in the following section.
We should add however that the UV model, theory A, is asymptotically free only for the SU group, but theory D is IR-free
for all gauge factors
with the beta functions being
$b^{\tiny (A)}_{\mbox{\tiny $\SU{\Nc}$}}=60$, $b^{\tiny (A)}_{\mbox{\tiny $\Sp{{M}}$}}=-7$,
$b^{\tiny (D)}_{\mbox{\tiny $\SU{\nct}$}}=-17$, $b^{\tiny (D)}_{\mbox{\tiny $\Sp{{m}}$}}=-1$.

\end{itemize}

\section{Supersymmetry breaking and the Sp sector}

Here we would like to comment on how the Sp factors which arose naturally in our approach as the mechanism for
deconfining antisymmetric matter fields of the $\SU{5}$ GUT and thus enabled us to construct its UV Seiberg dual(s),
can also trigger supersymmetry breaking in the low-energy effective SU GUT.\footnote{We also note that the Sp dynamics
is ultimately responsible for slowing down proton decay, as will be discussed later on.}
We shall investigate two complimentary scenarios for this to happen. In the first case we shall work with a theory D which contains
IR-free Sp factors. This allows one to carry out a weak coupling analysis of this theory in the IR which reproduces the
 ISS mechanism for SUSY-breaking in a long-lived metastable vacuum.
While rather satisfying conceptually, we have seen that actual examples (at least in the context of models considered in this paper)
are somewhat eccentric with large gauge groups and steep slopes. On the other hand, much simpler examples occur in the alternative
scenario, where Sp factors are S-confined in the IR. In these cases the SUSY-breaking potential of Sp factors
is complicated by the properties of the K\"ahler potential near the origin. This however does not rule out that SUSY breaking in the
S-confining case can occur \cite{ISS}. We now discuss both these possibilities.

\subsection{IR-free Sp dynamics}

Our main goal here is to demonstrate that Sp sectors can break SUSY in a calculable way in the infrared.
To this end we consider one of the `simpler' examples of the model with an IR-free Sp gauge groups.
Specifically, we take the example from the previous section with IR-free $\SU{9}\times\Sp{1}^3$ dual to an
$\SU{43}\times \Sp{4}^3$ GUT with $ \df=2$.

In the UV we have theory A, precisely as in table~\ref{deconf1} where $N=43$, $M=4$ and $\df=2$
and there is a single $Z^a$ field for each Sp factor.
This theory has an asymptotically free SU sector (with a rather steep slope)
$b^{\tiny (A)}_{\mbox{\tiny $\SU{\Nc}$}}=60$, and a non-asymptotically-free
Sp sectors with $b^{\tiny (A)}_{\mbox{\tiny $\Sp{{M}}$}}=-7$. We will tacitly assume that the latter
undergo further transformations before they reach their Landau poles.\footnote{Ideally it would be nice to confine these
Sp factors in the UV or decouple them all together. We shall not dwell on this feature of the UV theory for this example
and instead look at the IR physics.}

After KSS-dualising the SU gauge factor, we arrive at theory B in agreement with table~\ref{modelb}. As before, transition
from theory B to theory C is realised by integrating out heavy mesons, but now we treat mesons
$ \Sigma_{i a \bar{J}}= Y_a X^i \tilde{F}_{\bar{J}}$
as being light. In total we have one $y_a$ field, one light $z_a$ field (as before), plus four $\Sigma_a$ fields transforming
in the fundamental of $\Sp{M}$
which should be added
to table~\ref{modelcp} in theory C.
Finally we perform the Sp duality and flow to theory D. Its field content is given in table~\ref{modeld}
with $n=9$ and $m=1$, plus
there are additional magnetic quarks transforming as $\fund$ of $\Sp{1}$, and mesons which are singlets.
The new magnetic quarks are $\tilde{\Sigma}_a$ which are dual to our light $\Sigma_a$ quarks of $\Sp{4}$, and the new mesons include
a singlet $\eta = \Sigma \Sigma$ which is a four-by-four matrix in flavour space (four flavours corresponding to $i=0,1$ times
two from $ \bar{J}=1,\df$ in $ \Sigma_{i a \bar{J}}$.)
The superpotential of theory D then must contain additional terms describing dual-quark-meson-dual-quark interactions
and a linear meson term \cite{ISS}:
\be
W_{D\, {\rm rank}} \, = \,  \varepsilon^{ab} \tilde{\Sigma}_a \, \eta\,   \tilde{\Sigma}_b \,-\, \mu_{\rm ISS}^2 \,{\rm tr}\, (\eta \,J_4)
\label{Wdnew}
\ee
Here $\varepsilon^{ab}$ is the symplectic structure of magnetic $\Sp{1}$ gauge group, $J_4$ is the symplectic structure
of the unbroken flavour subgroup $\Sp{2}$, and $\mu_{\rm ISS}^2$ is the ISS parameter arising from the masses of
$\Sigma$ quarks of the electric dual theory C.
Because the number of Sp flavours (i.e. 4) is greater than the number of $\Sp{1}$ colours (i.e. 2), the $F$-terms arising
from differentiating $W_D$ in \eqref{Wdnew} with respect to $\eta$ cannot be all set to zero. This is precisely the
ISS rank condition which gives a SUSY-breaking vacuum. There is also a lower-lying SUSY-preserving vacuum which appears
parametrically far away in the field space due to the effect of a non-perturbatively generated superpotential contribution to $W_D$.
This implies that the SUSY-breaking vacuum caused by the rank condidtion is an exponentially long-lived metastable ground state \cite{ISS}.

We now need to couple the messengers $f$ and $\tilde{f}$ of the IR theory D to the SUSY-breaking $F_\eta$-terms so that supersymmetry
breaking is straightforwardly mediated to the Standard Model SU sector(s). To this end we introduce to the superpotential
of theory D the terms \be
W_{D\, {\rm mediation}} \, = \,  \kappa \,{\rm tr}\, (\eta \,J_4) \, \tilde{f} f \, +\, m_f \,\tilde{f} f\, ,
\label{Wdmed}
\ee
where the first term is the coupling between the `spurion' $\eta$ and the messengers (with $\kappa$ being a constant to be determined)
and the second term is the mass term for the messengers. We stress that $f$ and $\tilde{f}$ are not a new ingredient in theory D,
they have been present all the time as fundamental-anti-fundamental pairs of $\SU{n}$ matter fields, see table~\ref{modeld}; only the couplings
in \eqref{Wdmed} are new.

Our next step is to trace the origin of $W_{D\, {\rm mediation}}$ in the superpotential of the UV theories.
The most economical presentation is to simply write down the new terms in the superpotential of theory A and to follow them down to theory D.
We write
\be
W_{A\, {\rm mediation}} \, = \,  -\frac{1}{M_P^{2+i+j}}
\varphi_f\,{\rm tr}\, ((YX^i\tilde{F})\cdot(YX^j\tilde{F}) \,J_4) \, +\, m^2 \varphi_f \,+\, \varphi_f \,\tilde{F} F
\label{WAmed}
\ee
Here $\varphi_f$ is a singlet field (it can be though of as a meson of dual quarks $\tilde{f} f$), and the rest of the fields are familiar from
table~\ref{deconf1}. The first term is a higher-dimensional operator, for simplicity we shall take the lowest values of $i=0=j$
where no adjoints $X$ appear giving the operator of dimension-5, which is suppressed by two powers of the
`high scale' $M_P$. After the KSS-duality of the SU factor this flows to theory B
with the superpotential correction
\be
W_{B\, {\rm mediation}} \, = \,  -\frac{\Lambda_A^2}{M_P^{2}}
\varphi_f\,{\rm tr}\, (\Sigma\Sigma) \,J_4) \, +\, m^2 \varphi_f \,+\, \Lambda_A \,\varphi_f \,\hat{\phi}_F \, + \, \tilde{f}\, \hat{\phi}_F\, f
\label{WBmed}
\ee
The $\Sigma$ mesons of theory B appeared in the first term as $\Sigma= \Lambda_A^{(-1)}(YX^i\tilde{F})$,
and the electric quarks $F$, $\tilde{F}$ got replaced by their mesons $\hat{\phi}_F= \Lambda_A^{(-1)}\tilde{F} F$. The last term in
\eqref{WBmed} corresponds to the usual triple dual-quark-meson-dual-quark coupling of a magnetic theory.

In passing from theory B to theory C we integrate out heavy measons: in the current context this means integrating out
$\varphi_f$ and $\hat{\phi}_F$. Solving the classical equation for $\varphi_f$ we get
\be
\hat{\phi}_F \,=\, \frac{\Lambda_A}{M_P^{2}}\,{\rm tr}\, (\Sigma\Sigma) \,J_4) \, -\, \frac{m^2}{\Lambda_A}
\ee
and substituting this to \eqref{WBmed} one recovers the superpotential of theory C
\be
W_{C\, {\rm mediation}} \, = \,
\frac{\Lambda_A}{M_P^{2}}\,{\rm tr}\, (\Sigma\Sigma) \,J_4) \,\tilde{f} f \, -\, \frac{m^2}{\Lambda_A} \,\tilde{f} f
\label{WCmed}
\ee
which upon Seiberg-dualising the Sp factor, finally gives in theory D:
\be
W_{D\, {\rm mediation}} \, = \,
\frac{\Lambda_A \Lambda_C}{M_P^{2}}\,{\rm tr}\, (\eta \,J_4) \,\tilde{f} f \, -\, \frac{m^2}{\Lambda_A} \,\tilde{f} f\, .
\label{WDmed}
\ee
This is equivalent to the desired gauge mediation superpotential in \eqref{Wdmed} where
$\kappa = (\Lambda_A \Lambda_C)/(M_P^2) \ll 1$ and $m_f = - m^2/\Lambda_A$.
Having recovered the superpotential \eqref{Wdmed}  or \eqref{WDmed}, the gauge mediation proceeds in the
expected fashion. In fact, this picture corresponds to the ordinary gauge mediation scenario in the context of the ISS
metastability, whose phenomenological consequences were already studied in Ref.~\cite{MN}. In particular the smallness
of the coupling $\kappa$ in the IR theory is welcome as it breaks an approximate $\U{1}_R$-symmetry thus allowing for the
generation of gaugino masses, but at the same time not destabilizing the ISS SUSY breaking vacuum \cite{MN,Intriligator:2007py,Abel:2007nr}.

\subsection{S-confining Sp in the infrared}

So far we have established how SUSY-breaking and mediation can occur in a calculable model
provided by the Sp gauge theory when it is in the IR-free phase (in theory D). It is however much easier to
construct scores of minimal and less-eccentric models that have an asymptotically free UV phase in all gauge factors,
but which do not exhibit IR-free Sp factors, instead having them
S-confined. In the Examples section above we presented a few such models where the Sp factors completely disappear
in the IR leaving behind a minimal $\SU{5}$ GUT.
As far as the SUSY-breaking is concerned, the loss of an IR free Sp theory makes it more difficult to treat.
There are essentially two possibilities: either SUSY is broken in a completely separate hidden sector (and then communicated to
our GUT through $f$ and $\tilde{f}$ messengers), or it is still the Sp factors which are responsible for SUSY-breaking
albeit in a `non-calculable' way. Let us discuss this possibility in more detail.

As we have said the transition from theory C to theory D involving the Seiberg duality $\Sp{M} \to \Sp{m}$, now gives an
empty magnetic theory, i.e. $m=0$, corresponding to the S-confinement of Sp factors in the IR. The empty magnetic
Sp group does not allow for magnetic Sp-quarks, nor for the usual cubic couplings of magnetic quarks and mesons.
Instead one only has a linear potential in the meson field as well as the dynamically generated  contribution
which we discussed earlier (cf. \eqref{Wdyn2}):
\be
W_{D'}  \ni\, {\rm tr} (m \, {\rm meson}) \,-\, \frac{{\rm Pf}({\rm meson})}{\Lambda_{C}^{n-3}}\, ,
\label{Wdyn33}
\ee
The non-perturbative superpotential is important at large values of meson VEVs, whereas the linear superpotential
is important near the origin. In the absence of cubic couplings, it is the only term in the superpotential which
potentially can break supersymmetry near the origin. As ISS have already noted
\cite{ISS},
it is the K\"ahler potential  which then determines whether supersymmetry is broken near the origin by these S-confined Sp models.
This is clearly not a `calculable' SUSY-breaking scenario. However if we assume that it can or does occur,
we end up with a very appealing picture with Sp factors not only allowing GUT Seiberg duals to exist and to slow
down proton decay, but also accounting for the SUSY-breaking hidden sector automatically.

\section{Doublet-triplet splitting}

One interesting question in the context of dual unification is what happens to the
doublet-triplet splitting of the higgs: i.e. the fact that in the minimal $\SU{5}$ magnetic theory we know that
the $\SU{2}_L$ part of the higgs multiplets must remain light, while the higgs triplets
should be given masses of order $M_{GUT}$. This is a well known fine-tuning problem which 
we cannot hope to solve here. However it is particularly important since splitting the higgs is of course vital to getting 
the correct unification, so this tuning should have a meaning in both the electric and magnetic descriptions. 
In this section we briefly address this point: in short, we will see that  the tuning of couplings in the magnetic description 
becomes a tuning of VEVs in the electric description, and vice-versa (i.e. small masses $\leftrightarrow$ small VEVs).

\subsection{Spontaneous breaking of the GUT symmetry}

We can examine the question of multiplet splitting in full generality by considering a mass splitting
in the electric $\SU{N}$ theory of standard KSS introduced earlier.
First we recap the GUT breaking in more detail for the $k=2$ case. We will set $s_0$=1, so that
$W=Tr(\frac{X^{3}}{3}+m_X\frac{X^{2}}{2}+\lambda X)$. The eigenvalues
are \begin{equation}
X_{\pm}=\frac{-m_X\pm\sqrt{m_X^{2}-4\lambda}}{2}\end{equation}
and the condition $Tr(X)=0$ fixes \begin{equation}
\lambda=-m_X^{2}\frac{r_{+}r_{-}}{(r_{+}-r_{-})^{2}}.\end{equation}
If $m_X>0$ (by assumption) there must be more of the $X_{+}$ eigenvalues,
and hence $r_{+}>r_{-}$ and \begin{eqnarray}
\sqrt{m_X^{2}-4\lambda} & = & m_X\frac{(r_{+}+r_{-})}{(r_{+}-r_{-})}.\\
X_{\pm} & = & \pm m_X\frac{r_{\mp}}{r_{+}-r_{-}},\\
X_{s\pm} & = & \pm\frac{m_X}{2}\frac{(r_{+}+r_{-})}{(r_{+}-r_{-})},\end{eqnarray}
where we have defined
new fields $X_{s}$ such that $
X_{s}=X+\frac{m_X}{2}\mathbf{1}.$
Then\be
W\,=\,\frac{X^{3}}{3}+m_X\frac{X^{2}}{2}+\lambda X  =  \frac{X_{s}^{3}}{3}+(\lambda-m_X^{2}/4)X_{s}\, .\ee
The masses of e.g. the fermions (supersymmetry is never broken by
assumption) are \be
W_{XX}=2X+m_X=\left\{ \begin{array}{cc}
m_X & \,\,;\,\, X\equiv X_{ij}\\
\pm m_X\frac{N_{c}}{r_{+}-r_{-}} & \,\,;\,\, X\equiv X_{ii}\end{array}\right.\ee
Note that $Tr(X_{s})=\frac{m_X}{2}\Nc$ and $X_s$ is not traceless.

The mesons $\Phi_{j=0\ldots k-1}$ are defined as before, and for ease of reading we repeat the superpotential in the magnetic theory:
\begin{equation}
\wmg=\sum_{i=0}^{k-1}\frac{-{t}_{i}}{k+1-i}Tr(x_{s}^{k+1-i})+\frac{1}{\mu^{2}}\sum_{l=0}^{k-1}t_{l}\sum_{j=1}^{k-l}\Phi_{j-1}\tilde{q}x_{s}^{k-j-l}q\label{eq:wmag3}\end{equation}
where  $t$ are coefficients in the $X_s$ (and $x_s$) basis. According to
KSS this then gives the magnetic superpotential \begin{equation}
W_{mg}=\,-\frac{x^{3}}{3}+{m}_x\frac{x^{2}}{2}+\bar{\lambda}x\end{equation}
where \begin{equation}
{m}_x=\frac{\Nc}{\nct}m_X\, .\end{equation}
We can check this is consistent. Indeed,
the eigenvalues are \begin{equation}
x_{\pm}=\frac{{m}_x\mp\sqrt{{m}_x^{2}+4\bar{\lambda}}}{2}\end{equation}
and the condition $Tr(x)=0$ fixes \begin{equation}
\bar{\lambda}={m}_x^{2}\frac{\bar{r}_{+}\bar{r}_{-}}{(\bar{r}_{+}-\bar{r}_{-})^{2}}\end{equation}
and then (since $\bar{r}_{-}>\bar{r}_{+}$) we have \begin{eqnarray*}
x_{\pm} & = & \pm{m}_x\frac{\bar{r}_{\pm}}{\bar{r}_{+}-\bar{r}_{-}}\\
\sqrt{{m}_x^{2}+4\bar{\lambda}} & = & {m}_x\frac{(\bar{r}_{+}+\bar{r}_{-})}{(\bar{r}_{-}-\bar{r}_{+})},\\
x_{s\pm} & = & \pm\frac{{m}_x}{2}\frac{(\bar{r}_{+}+\bar{r}_{-})}{(\bar{r}_{-}-\bar{r}_{+})}\end{eqnarray*}
The masses of e.g. the fermions is \[
W_{xx}=-2x+{m}_x=\left\{ \begin{array}{cc}
{m}_x & \,\,;\,\, x\equiv x_{ij}\\
\pm{m}_x\frac{n}{\bar{r}_{+}-\bar{r}_{-}} & \,\,;\,\, x\equiv x_{ii}\end{array}\right.\]
Note that the masses of the diagonal components are the same as in
the electric theory: \[
\pm{m}_x\nct\frac{1}{\bar{r}_{+}-\bar{r}_{-}}=\pm m\Nc\frac{1}{r_{-}-r_{+}}\]
This mapping is consistent because in the magnetic theory the shift we would make to remove the $x^{2}$ term is $x_{s}=x-\frac{\bar{m}}{2}\mathbf{1}$
and then \[
\wmg\supset -\frac{x_{s}^{3}}{3}+(\bar{\lambda}+{m}_x^{2}/4)x_{s}\]
and one can check that \begin{eqnarray}
(\bar{\lambda}+{m}_x^{2}/4) & = & \frac{({m}_xn)^{2}}{4(\bar{r}_{+}-\bar{r}_{-})^{2}}\nonumber \\
 & = & (m_X^{2}/4-\lambda),\end{eqnarray}
so that \be
x_{s\pm}=X_{s\pm}\, ,\ee
as required.

\subsection{Splitting the higgs}

Now we deform the theory with a doublet/triplet mass term for the
last flavour of fundamental which we will label $H$ (for higgs). We
allow finite masses for everyone:\be
W\supset\frac{m_{h}}{2}\left(\frac{r_{+}+r_{-}}{r_{+}-r_{-}}\right)\tilde{H}H+\tilde{H}X_{s}H.\ee
One could of course use $\tilde{H}XH$ instead, however in this notation
the second term will correspond directly to the linear meson term in the magnetic theory.
The $X_{s}$ VEVs have split the gauge group so there are two higgs
multiplets for each with different masses. They are \be
\frac{m_{h}}{2}\left(\frac{r_{+}+r_{-}}{r_{+}-r_{-}}\right)+X_{s\pm}  =  \frac{(m_{h}\pm m_X)}{2}\left(\frac{r_{+}+r_{-}}{r_{+}-r_{-}}\right),
\ee
giving $H$ masses. To split the multiplet in the conventional manner  we
define a (finely tuned) small parameter
\be
\epsilon=m_{h}-m_X\, .\ee
This gives $r_{-}$ small masses for the $\SU{r_{-}}$ part of the  multiplet, 
so we are doing $r_-$plet--$r_+$plet mass splitting.
Note that since the higgses do not get VEVs the VEVs of the adoints
is unchanged.

What happens in the magnetic theory? In that description
the mass terms turn into linear meson terms in the superpotential;\begin{eqnarray}
\wmg & = & -\frac{x_{s}^{3}}{3}+(\bar{\lambda}+{m}_x^{2}/4)x_{s}+\frac{m_{h}}{2}\left(\frac{r_{+}+r_{-}}{r_{+}-r_{-}}\right)\Phi_{0}^{(h)}+\Phi_{1}^{(h)}\\
 &  & \,\,\,\,\,\,+\frac{1}{\mu^{2}}\left(\Phi_{0}\tilde{q}x_{s}q+\Phi_{1}\tilde{q}q\right).\end{eqnarray}
We are taking the $N_{f}$'th
generation of fundamentals to be the higgs. Therefore we have by the $x_{s}$
and $\Phi_{0,1}^{(h)}$ equations of motion that \begin{eqnarray}
0 & = & \tilde{h}x_{s}h+\mu^{2}\frac{m_{h}}{2}\left(\frac{r_{+}+r_{-}}{r_{+}-r_{-}}\right)\\
0 & = & \tilde{h}h+\mu^{2}\\
0 & = & -x_{s}^{2}+(\bar{\lambda}+{m}_x^{2}/4)\end{eqnarray}
with contraction over colour indices implied.
We can by making suitable gauge rotations (and imposing $D$-flatness)
choose a basis in which there are two non-zero entries in each of
$h$, $\tilde{h}$, one of them the last entry in the $\bar{r}_{+}$
group, and one the last entry in the $\bar{r}_{-}$ group. The two
equations above then give \begin{eqnarray}
h_{+}^{2}-h_{-}^{2} & = & \mu^{2}\frac{m_{h}}{m_X}\nonumber \\
h_{+}^{2}+h_{-}^{2} & = & \mu^{2}\end{eqnarray}
and solving gives\begin{equation}
h_{\pm}^{2}=\mu^{2}\frac{(m_{h}\pm m_X)}{2m_X}.\end{equation}

Thus we see that generally the picture is as follows; keeping the $\SU{r_{-}}$ higgs
light in the electric theory means that we can integrate out a single
heavy (mass order $m_X$) $\SU{r_{+}}$ quark (i.e. the higgs triplet).
Choosing $m_X=m_{h}$ gives $r_{-}-$plet/$r_{+}$-plet splitting:
the effective electric theory is therefore $\SU{r_{+}}\times \SU{r_{-}}$
with $N_{f}-1$ and $N_{f}$ flavours of fundamental respectively. The magnetic
theory is correspondingly higgsed in only the $\SU{\bar{r}_{+}}$
group as $\SU{\bar{r}_{+}}\rightarrow \SU{\bar{r}_{+}-1}$,
as required since the electric $\SU{r_+}$ has lost a flavour and
by Seiberg duality we know that we should have
$\bar{r}_{+}=N_{f}-1-r_{+}$. Of course the reverse
can be arranged by choosing $m_{h}+m_X=0$, and generic values give
the single integrated out quark as in KSS. Likewise we may arrange
to have split masses in the magnetic theory and split VEVs in the electric.

\subsection{Generalisation}

Not surprisingly this picture can be shown to hold in the most
general set up. Consider
a general value of $k$ and an electric superpotential \be
\wel= g(X_{s})+\tilde{H}f(X_{s})H\ee
where
\be
f(X_{s})  =  \sum_{j=1}^{k}c_{j}X_{s}^{j-1}
\ee
is a general $k-1$'th order polynomial. Assume that $g(X_{s})$ gets
$k$ independent eigenvalues labelled $X_{s,i}$ in the usual manner.
Then a vanishing mass term corresponds to one of the $X_{s,i}$ coinciding
with one of the roots of $f(X_{s})$.

The magnetic theory is arranged so that $x_{s,i}=X_{s,i}$ which is
our only assumption. The magnetic mesons are $\Phi_{j}^{(h)}=\tilde{H}X_{s}^{j}H$,
so the magnetic superpotential is \be
\wmg
=-g(x_{s})+
\sum_{j=1}^{k}c_{j}\Phi_{j-1}^{(h)}+\frac{1}{\mu^{2}}\sum_{l=0}^{k-1}t_{l}\sum_{j=1}^{k-1}\Phi_{j-1}^{(h)}\tilde{h}x_{s}^{k-j-l}h.\ee
This can be rewritten\be
\wmg
=-g(x_{s})+\sum_{j=1}^{k}\Phi_{j-1}^{(h)}\left(c_{j}+\frac{1}{\mu^{2}}\sum_{l=0}^{k-j}t_{l}\tilde{h}x_{s}^{k-j-l}h\right),\ee
so the $\Phi_{j}^{(h)}$ equations of motion give
\be
c_{j}+\frac{1}{\mu^{2}}\tilde{h}\left(\sum_{l=0}^{k-j}t_{l}x_{s}^{k-j-l}\right)h=0\,\,\,\,\forall j=1\ldots k\ee
Again we can choose a basis in which the higgs VEVs have a single
nonzero entry in the first element of each of the $k$ sub-groups;
calling these VEVs $h_{i=1\ldots k}$ and setting $\tilde{h}=-h$
to zero the $D$-terms (the phase corresponds to a $\U{1}_{B}$ rotation
- this group being broken by the deformations), we get
\be
c_{j}(x_{s,i})^{j-1}=\frac{1}{\mu^{2}}\left(\sum_{l=0}^{k-j}t_{l}(x_{s,i})^{k-j-1}\right)h_{i}^{2}\,\,\,\,\forall j=1\ldots k
\ee
Summing over $j$ we find \be
f(x_{s,i})=\frac{1}{\mu^{2}}g''(x_{s,i})\, h_{i}^{2}\, ,\ee
which, since the VEVs in the dual theories match, is the same as\be
f(X_{s,i})=\frac{1}{\mu^{2}}g''(X_{s,i})\, h_{i}^{2}.\ee
In other words whenever $X_{s,i}$ coincides with a root of $f(X)$
giving a massless higgs in the electric theory, either $h_{i}=0$
in the magnetic theory, or the corresponding adjoint field is massless.
This holds for any number of light higgses in a multiply split multiplet.

\section{Proton decay in dualified $\SU{5}$}

In Ref.~\cite{AK:Dualification}, it was argued that in dual unification, even if
the Landau pole is only slightly below $M_{GUT}$, the proton decay
is hugely suppressed. In that paper the arguments were presented
in general terms because of the absence of a convincing dual for
$\SU{5}$ GUTs. Now that we have one, it is worth revisiting the
proton decay issue to see explicitly how the suppression happens.

\begin{figure}
\begin{centering}
\includegraphics[scale=0.6]{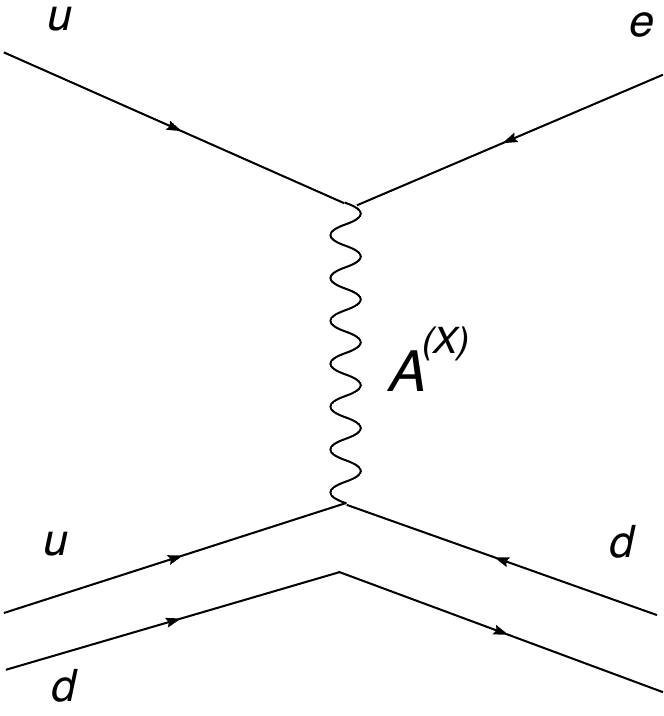}
\hspace{1cm}
\includegraphics[scale=0.7]{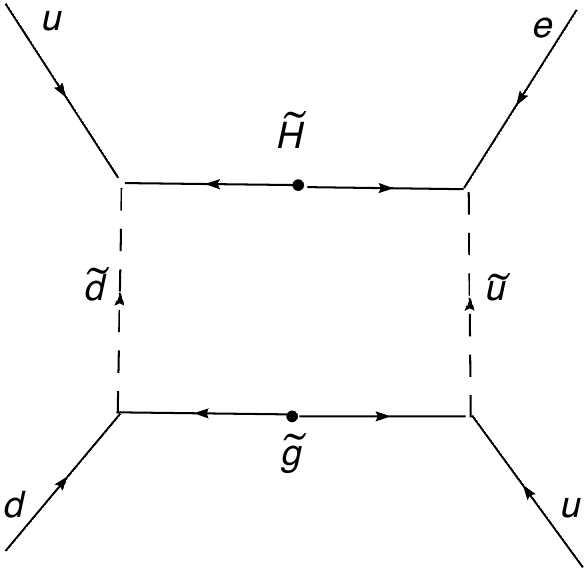}
\par\end{centering}
\caption{\it \label{pdecay} Proton decay in simple $SU(5)$ SUSY
GUTs generated by dimension 6 and dimension 5 operators
respectively.}
\end{figure}

First let us give the general arguments again. The proton decay in
GUTs is due to the presence of GUT bosons and heavy
coloured triplets (for reviews see
Refs.~\cite{Raby:2008zz,decayreview,decayreview2}). If one assumes
simple unification in SUSY $\SU{5}$ at the usual scale
$M_{GUT}\approx 2\times 10^{16}$~GeV, the resulting lifetime of
the proton is shorter than the present experimental bound of
$\tau_p \gtrsim 6.6 \times 10^{33}yrs$, and simple unification
seems to be ruled out. Indeed in supersymmetric $SU(5)$ the proton
is able to decay via two types of diagram. The first, which it
shares with {\em non}-supersymmetric $\SU{5}$, is
 gauge boson exchange, as in figure \ref{pdecay}a: it generates dimension 6 operators
in the effective potential such as
\begin{equation}
{\cal L}_{eff} \supset \frac{g^2}{2M_{GUT}^2}
\varepsilon_{ijk}\varepsilon_{ab}( \bar{q}_{aj}  \gamma_\mu
{u}^c_{k} ) (\bar{q}_{ib}\gamma_\mu e^+)\,
\end{equation}
allowing the proton to decay via processes such as $p\rightarrow
\pi^0 e^+ $. These process are not normally considered dangerous
since by themselves they would give a lifetime of $\tau_p \sim
10^{34-38}$ yrs which can still be accommodated.
In supersymmetric theories however the dominant decays are via
dimension 5 operators that contribute at one-loop due to the
presence of higgs triplets, $\tilde{h}_T\equiv {\bf \bar{3}}$ and
$h_T\equiv {\bf 3}$, that couple via the Yukawa couplings of the
MSSM: \be W\supset \frac{\lambda_u}{4} \varepsilon^{(5)} aah +
\lambda_d \tilde{q} a \tilde{h}
    \supset
\lambda_u u^c_i e^c { h}_{T\,i} + \lambda_d \,\varepsilon^{(3)}\,
d^c u^c {\tilde h}_{T} \, , \ee and similar for left handed
fields, where $\lambda_u$ and $\lambda_d$ are the Yukawa couplings
of the MSSM. These give rise via figure \ref{pdecay}b to the most
dangerous operators, for example \be {\cal L}_{eff} \supset
\frac{g^2 \lambda_u \lambda_d}{16 \pi^2 M_{SUSY}  M_{GUT}}
\varepsilon^{(3)}_{ijk}( {u}^c_{i} e^c ) ( u^c_{j} d^c_{k} )  \, .
\ee The resulting proton lifetimes are typically less than the
measured limits and it has been known for some time that this is
enough to rule out minimal SUSY $\SU{5}$
\cite{Goto:1998qg,Murayama:2001ur}.

\begin{figure}
\begin{centering}
\includegraphics[scale=0.8]{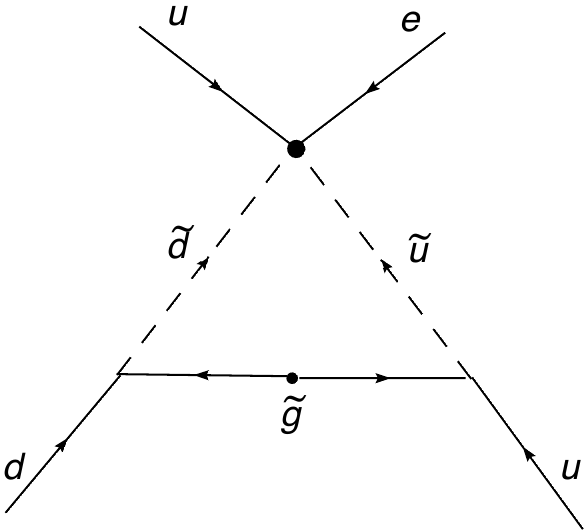}
\par\end{centering}
\caption{\it \label{pdecayapp} Approximation to figure
\ref{pdecay}b in which the dimension 5 operator is an effective
operator in the superpotential.}
\end{figure}

Note that in this estimate, thanks to the non-renormalization
theorem, the one loop integral is dominated by the low momentum
region $k\lesssim M_{SUSY}$, and so $M_{SUSY}$ appears in the
denominator. Hence the diagram can be approximated by first
integrating out the heavy higgs triplet, with the momentum in the
rest of the loop being around the weak scale. When the higgs
triplet is integrated out it generates a set of baryon number
violating non-renormalizable terms in the effective theory, \be
\label{weff} W_{eff} \supset c_R \, \varepsilon^{(3)}   {u}^c u^c
d^c e^c +c_L \, \varepsilon^{(3)} q\, q\, q\, l  \, , \ee where
the typical coefficients in standard $\SU{5}$ are \be c_{R}\approx
c_{L} \approx \frac{\lambda_u \lambda_d}{M_{GUT}} \, . \ee These
vertices replace the higgs propagator in figure \ref{pdecay}b with
the corresponding 4-point vertex as in figure \ref{pdecayapp},
which can be evaluated entirely within the low energy theory.

Now let us turn to a dual unified theory. The first point we
should stress is that the operators in \eqref{weff} descend from a
holomorphic baryon operator of $\SU{5}$, namely \be \label{weff2}
W_{eff} \supset c_5 \, \varepsilon^{(5)}   a a (a \tilde{q})  \, ,
\ee with \be \label{dooper} c_{5} \approx \frac{\lambda_u
\lambda_d}{M_{GUT}} \, , \ee where $(a \tilde{q})$ are contracted
to give the 5th fundamental index. The diagram in figure
\ref{pdecay}b can now not be done directly because some of the
scales (i.e. those on the higgs propagator) are GUT size, while 
the diagram is dominated by weak scale momenta. However
we may readily approximate the diagram as we did in figure
\ref{pdecayapp}; that is, first we integrate out the GUT mass
states in the electric theory, then we map them to the low energy
magnetic theory to find the effective superpotential $W_{eff}$,
specifically its coefficients $c_5$. Once we have derived
$W_{eff}$, the proton decay rate can be evaluated exactly as
above, with $W_{eff}$ simply providing an effective 4 point
coupling as in figure \ref{pdecayapp}. This procedure is
straightforward as long as we know how to map the relevant
electric operators to magnetic ones. Fortunately this matching of
baryons is known and is indeed an integral part of the Seiberg
duality as described in section 2.

Before providing the exact mapping, let us discuss heuristically
what should happen. We begin first in the electric theory at the
GUT scale by integrating out heavy $M_{GUT}$ mass states. This is
entirely perturbative (and indeed tree-level) and the theory has
no idea that it is about to get strongly coupled: hence the only
dimensionful parameter that can enter into the result is $M_{GUT}$
itself. Thus, by dimensions, the electric superpotential after
integrating out all the heavy states will contain operators such
as \be \label{weff3} W^{(el)}_{eff} \supset
\frac{\lambda^{N-2}}{M_{GUT}^{N-3}} \, \varepsilon^{(N)}  Q^N  \,
. \ee Here the $\lambda$ indicates a generic coupling between a
heavy GUT state and two light ones. In a theory (like our theory
A) that has only quarks (i.e. no antisymmetrics) a tree-level
diagram with $N$ external quark legs has $N-2$ such vertices. One
may also have (as we shall see) diagrams with extra GUT adjoints,
in which case extra couplings would appear. We now run the theory
down to the scale at which it becomes strongly coupled and pass to
the magnetic description. In going between the two theories we can
map the baryons as described in section 2. The matching is of the
form \be \varepsilon^{(N)}  Q^N \leftrightarrow \Lambda ^{N-5}
\varepsilon^{(5)} q^5\, . \ee Note that the only dimensionful
parameter that can appear here is the dynamical scale $\Lambda$.
(Indeed the matching must be the same as for the undeformed
theory with only the $X^3$ term, and in this theory $M_{GUT}$
doesn't even appear.)
Hence the holomorphic baryon operator appearing in the magnetic
theory is \be \label{weff4} W_{eff} \supset c_5 \,
\varepsilon^{(5)}  q^{(5)}  \, , \ee with \be c_{5} \approx
\frac{\lambda^{N-2} }{M_{GUT}} \left(
\frac{\Lambda}{M_{GUT}}\right)  ^{N-5} \, . \ee This coefficient
is of course tiny compared to the $c_5$ one would expect in the
conventional SUSY $\SU{5}$, i.e. that in \eqref{dooper}: it is
clear that even with a moderately low Landau pole and moderately
small couplings, the proton decay is enormously suppressed.

Let's turn to a realistic example. We will consider the
$\SU{11}\times \Sp{1}^3$ model in which the Sp gauge groups
eventually confine in the IR. The simplest procedure is first to
track the offending operator back through the chain of dualities
from theory D to theory A. Once we have identified it in theory A
we can work forwards again to determine the coefficient $c_5$ in
the effective superpotential of theory D.

Before we start we should briefly mention the dynamical scales: in
order to simplify the discussion we will set all the dynamical
scales in the GUT to be degenerate \be\Lambda_D\sim
\Lambda_C\sim \Lambda_B\sim \mu_B\sim \Lambda\ll M_{GUT}\, ,\ee and
assign all the couplings of the adjoint superpotential (i.e. the
$t_i$) values of order one in $M_{GUT}$ units. 
Now, we begin with the baryon \be b^{(D)} = \varepsilon^{(5)}   a
a (a \tilde{q}) \, . \ee The first step is to deconfine to theory
C and open up the antisymmetrics. The baryon in theory C is
therefore \be b^{(C)} = \Lambda^3\, \varepsilon^{(5)}   y.y\,
y.y\,   y. (y\tilde{q}) \, \, , \ee where the dots indicate Sp
contraction and the bracket indicates an $\SU{5}$ contraction.
Note that the story deviates slightly from the heuristic version
given above: this is an $\SU{5}$ baryon times by an $\SU{5}$ meson
operator so we are going to have to treat each individually. First
consider the Sp valued baryon $\varepsilon^{(5)}   y.y\, y.y\,
y$ when we perform the SU duality. This baryon operator clearly
won't change when we integrate in some mesons to get to theory B,
so for this part we can proceed directly to do the SU duality to
theory A. The matching of such operators is complimentary in the
manner described in section 2. In fact, following that discussion,
the general expression for baryon matching is \be
\varepsilon^{(N)} \,  F^{{m}_0} (XF)^{{m}_1} Y^{{n}_0}
(XY)^{{n}_1}\leftrightarrow
\Lambda^{N-n+m_1-\bar{m}_1+n_1-\bar{n}_1} \varepsilon^{(n)} \,
f^{\bar{m}_0} (xf)^{\bar{m}_1} y^{\bar{n}_0} (xy)^{\bar{n}_1} \ee
where \ba
m_0+m_1 + n_0 + n_1 & = & \Nc \nonumber \\
\bar{m}_i &=& \df - m_i \nonumber \\
\bar{n}_i &=& 6 M  - n_i \, . \ea Note that with this definition
we have $\bar{m}_0+\bar{m}_1 + \bar{n}_0 + \bar{n}_1 =2 (6M+\df)
-N=n$ as required for the Levi-Cevita contractions.  For the
baryon operator of interest here we have $\bar{n}_0=5$ with
$\bar{n}_1=\bar{m}_0=\bar{m}_1=0$, giving \be \varepsilon^{(11)}
\,  F^{2} (XF)^{2} Y^{} (XY)^{6}
 \leftrightarrow
\Lambda^{14} \, \varepsilon^{(5)}   y.y\, y.y\,   y\, \ee

Now for the other factor: in theory C this operator is to be
contracted with the fundamental Sp meson $(y \tilde{q})$. In
theory B this corresponds to Sp contraction with a heavy
${\Phi}_0$ (and/or a $\Phi_1$) meson. To see this let us add the
relevant term to $W^B$; the important terms in the superpotential
are \be W^{(B)} \supset  \bar{\lambda} \Phi_0 \chi_1 + \Phi_0
\tilde{h}x_s y + \chi_1 \tilde{q} y - \hat{O}{{\Phi}_0} \ee where
$\hat{O}$ is the rest of the operator, namely $\varepsilon^{(5)}
\, y.y \, y.y \, y\,$. As before, the first term is a Dirac mass
and we can integrate out $\Phi_0$ and $\chi_1$ leaving the same
operators in $W^{(C)}$ as before but now with the required
additional piece, \be W^{(C)} \supset
 \hat{O} \tilde{q} y\, .
\ee

Finally this identification allows us to map the whole operator
back to theory A since $\Phi_0 \leftrightarrow Y \tilde{Q}$; \be
(\varepsilon^{(11)} \,  F^{2} (XF)^{2} Y^{} (XY)^{6})  (Y
\tilde{Q}) \leftrightarrow  \Lambda^{17} \varepsilon^{(5)}   a a
(a \tilde{q}) \, . \ee The operator on the right is the effective
operator that we can put in to figure \ref{pdecayapp} in order to
calculate the proton decay. The operator on the left is its
equivalent that has to be generated at the GUT scale in theory A.
In other words one would expect that in the full GUT a suitable
tree-level diagram would produce a term (setting all the couplings
to 1) \be W^{(A)} \supset \,  \frac{1}{M_{GUT}^{18}}  (
\varepsilon^{(11)} F^{2} (XF)^{2} Y^{} (XY)^{6} ) (Y \tilde{Q})\,
, \ee and that at low energies this corresponds to \be W^{(D)}
\supset  \,  \left( \frac{\Lambda}{M_{GUT}}\right) ^{17} \frac{
\varepsilon^{(5)} a a (a \tilde{q})}{M_{GUT}}\, . \ee One might
wonder what can be said about the less dangerous dimension 6
operators. Since these are not holomorphic the answer is
unfortunately not much. Nevertheless it seems inconceivable that a
similar suppression would not take place for them too, although we
have no proof.

\section{Conclusions}

We have examined the possibility that the 
MSSM is a unified $\SU{5}$ theory that encounters a Landau pole below 
the GUT scale driven by messenger fields. 
We presented a number of possible electric Seiberg dual 
GUTs that may provide a consistent UV completion to the 
minimal $\SU{5}$ model: particularly nice examples are the
asymptotically free $\SU{11}\times \Sp{1}^3$ and $\SU{9}\times \Sp{1}^3$ models which are both nonchiral. 
In the magnetic theory the Sp groups become confining and generate the thee generations of antisymmetrics of the 
minimal $\SU{5}$ GUT as well as the higgs fundamental as composite fields.  They also generate the 
up-quark Yukawas nonperturbatively.

 As was recently pointed out in
Ref.~\cite{AK:Dualification}, unification predictions would be
transmitted across such a duality, but proton decay would be
negligible. We demonstrated that the dangerous dimension 5 operators
are indeed suppressed by many orders of magnitude (in the $\SU{11}\times \Sp{1}^3$ model for example we find 
a factor $(\Lambda/M_{GUT})^{17}$ where $\Lambda$ is the typical Landau pole scale). We would expect similar suppression for 
the dimension 6 operators.

Finally we also
discussed how the gauge-mediated supersymmetry breaking can be
incorporated into an overall unified picture using metastable
supersymmetry breaking. In the case where the Sp groups confine, the 
supersymmetry breaking would be incalculable, but the idea that the 
Sp groups are responsible for generating the antisymmetrics of the Standard Model, 
for generating the Yukawa couplings {\em and} for SUSY breaking is attractive and very minimal, 
and we think it deserves further study.

\subsection*{Acknowledgements}
We would like to thank J. Barnard, C. Csaki, J. Jaeckel, Y. Shirman and J. Terning for discussions, helpful comments and observations.
Both authors are supported by Leverhulme Trust research fellowships. 


\begin{thebibliography}{99}



\bibitem{AK:Dualification}
S.~Abel and V.~V.~Khoze,
  ``Direct Mediation, Duality and Unification,''
  JHEP {\bf 0811} (2008) 024
  [arXiv:0809.5262 [hep-ph]].

\bibitem{Abel:2007jx}
  S.~Abel, C.~Durnford, J.~Jaeckel and V.~V.~Khoze,
  ``Dynamical breaking of $U(1)_{R}$ and supersymmetry in a metastable vacuum,''
  Phys.\ Lett.\  B {\bf 661} (2008) 201
  [arXiv:0707.2958 [hep-ph]].


\bibitem{ISS}
  K.~A.~Intriligator, N.~Seiberg and D.~Shih,
  ``Dynamical SUSY breaking in meta-stable vacua,''
  JHEP {\bf 0604} (2006) 021
  [arXiv:hep-th/0602239].


\bibitem{S:Duality}
{N. Seiberg},
\newblock {``Electric-magnetic duality in supersymmetric non-abelian gauge
  theories"},
\newblock { {Nucl. Phys. B}} {\bf 435(1)},  129--146 (1995), [arXiv:hep-th/9411149].


\bibitem{Intriligator:1995au}
  For a review see K.~A.~Intriligator and N.~Seiberg,
  ``Lectures on supersymmetric gauge theories and electric-magnetic  duality,''
  Nucl.\ Phys.\ Proc.\ Suppl.\  {\bf 45BC}, 1 (1996)
  [arXiv:hep-th/9509066].

\bibitem{Strassler:1996ua}
 M.~J.~Strassler,
  ``Duality in supersymmetric field theory: General conceptual background and
  an application to real particle physics,''
{\tt http://www.slac.stanford.edu/spires/find/hep/www?irn=4969456}{SPIRES entry}
{\it Prepared for International Workshop on Perspectives of Strong Coupling Gauge Theories (SCGT 96), Nagoya, Japan, 13-16 Nov 1996}

\bibitem{K:Adj}
{D. Kutasov},
\newblock {``A comment on duality in $\mathcal{N}=1$ supersymmetric non-abelian
  gauge theories"},
\newblock { {Phys. Lett. B}} {\bf 351}, 230--234 (1995), [arXiv:hep-th/9503086].

\bibitem{KS:DKSS}
{D. Kutasov, A. Schwimmer},
\newblock {``On duality in supersymmetric Yang-Mills theory"}
\newblock { {Phys. \ Lett.\ B}} {\bf  354}, 315--321 (1995), [arXiv:hep-th/9505004].

\bibitem{KSS:DKSS}
{D. Kutasov, A. Schwimmer, N. Seiberg},
\newblock {``Chiral rings, singularity theory and electric-magnetic duality"},
\newblock {\em {Nucl. Phys. B}}{\bf 459}, 455--496 (1996), [arXiv:hep-th/9510222v1].


\bibitem{B:2Adj}
  J.~H.~Brodie,
  ``Duality in supersymmetric $\SU{N_c}$ gauge theory with two adjoint chiral
  superfields,''
  Nucl.\ Phys.\  B {\bf 478}, 123 (1996)
  [arXiv:hep-th/9605232].

\bibitem{BS:Theatre}
{J.H. Brodie, M.J.Strassler},
\newblock {``Patterns of duality in $\mathcal{N}=1$ SUSY gauge theories"},
\newblock { {Nucl. Phys. B}} {\bf 524}, 224--250 (1998), [arXiv:hep-th/9611197].

\bibitem{ILS:NewDualities}
{K. Intriligator, R.G. Leigh, M.J. Strassler},
\newblock {``New examples of duality in chiral and non-chiral supersymmetric
  gauge theories"},
\newblock { {Nucl. Phys. B}} {\bf 456}:567--621 (1995), [arXiv:hep-th/9506148].



\bibitem{Poppitz:1996wp}
  E.~Poppitz, Y.~Shadmi and S.~P.~Trivedi,
  ``Supersymmetry breaking and duality in SU(N) x SU(N-M) theories,''
  Phys.\ Lett.\  B {\bf 388}, 561 (1996)
  [arXiv:hep-th/9606184].

\bibitem{Poppitz:1996vh}
  E.~Poppitz, Y.~Shadmi and S.~P.~Trivedi,
  ``Duality and Exact Results in Product Group Theories,''
  Nucl.\ Phys.\  B {\bf 480}, 125 (1996)
  [arXiv:hep-th/9605113].

\bibitem{Klein:1998uc}
  M.~Klein,
  ``More confining N = 1 SUSY gauge theories from non-Abelian duality,''
  Nucl.\ Phys.\  B {\bf 553}, 155 (1999)
  [arXiv:hep-th/9812155].


\bibitem{Klein:2003wa}
  M.~Klein and S.~J.~Sin,
  ``On effective superpotentials and Kutasov duality,''
  JHEP {\bf 0310}, 050 (2003)
  [arXiv:hep-th/0309044].


\bibitem{singlets}
  S.~Abel and J.~Barnard,
  ``Electric/Magnetic Duality with Gauge Singlets,''
  JHEP {\bf 0905} (2009) 080
  [arXiv:0903.1313 [hep-th]].


\bibitem{Klebanov:2000hb}
I.~R.~Klebanov and M.~J.~Strassler,
  ``Supergravity and a confining gauge theory: Duality cascades and
  chiSB-resolution of naked singularities,''
  JHEP {\bf 0008}, 052 (2000)
  [arXiv:hep-th/0007191].

\bibitem{Berkooz:1997bb}
  M.~Berkooz, P.~L.~Cho, P.~Kraus and M.~J.~Strassler,
  ``Dual descriptions of SO(10) SUSY gauge theories with arbitrary numbers  of
  spinors and vectors,''
  Phys.\ Rev.\  D {\bf 56}, 7166 (1997)
  [arXiv:hep-th/9705003].

\bibitem{Berkooz:1995km}
  M.~Berkooz,
  ``The Dual of supersymmetric SU(2k) with an antisymmetric tensor and
  composite dualities,''
  Nucl.\ Phys.\  B {\bf 452}, 513 (1995)
  [arXiv:hep-th/9505067].

\bibitem{Pouliot:1995zc}
  P.~Pouliot,
  ``Chiral duals of nonchiral SUSY gauge theories,''
  Phys.\ Lett.\  B {\bf 359}, 108 (1995)
  [arXiv:hep-th/9507018].

\bibitem{Pouliot:1995me}
  P.~Pouliot,
  ``Duality in SUSY $SU(N)$ with an Antisymmetric Tensor,''
  Phys.\ Lett.\  B {\bf 367}, 151 (1996)
  [arXiv:hep-th/9510148].

\bibitem{Pouliot:1995sk}
  P.~Pouliot and M.~J.~Strassler,
  ``A Chiral $SU(N)$ Gauge Theory and its Non-Chiral $Spin(8)$ Dual,''
  Phys.\ Lett.\  B {\bf 370}, 76 (1996)
  [arXiv:hep-th/9510228].

\bibitem{Leigh:1997sj}
  R.~G.~Leigh, L.~Randall and R.~Rattazzi,
  ``Unity of supersymmetry breaking models,''
  Nucl.\ Phys.\  B {\bf 501} (1997) 375
  [arXiv:hep-ph/9704246].


\bibitem{Intriligator:1995ne}
  K.~A.~Intriligator and P.~Pouliot,
  ``Exact superpotentials, quantum vacua and duality in supersymmetric SP(N(c))
  gauge theories,''
  Phys.\ Lett.\  B {\bf 353}, 471 (1995)
  [arXiv:hep-th/9505006].

\bibitem{Affleck:1983rr}
  I.~Affleck, M.~Dine and N.~Seiberg,
  ``Supersymmetry Breaking By Instantons,''
  Phys.\ Rev.\ Lett.\  {\bf 51}, 1026 (1983).
  
\bibitem{Affleck:1983mk}
  I.~Affleck, M.~Dine and N.~Seiberg,
  ``Dynamical Supersymmetry Breaking In Supersymmetric QCD,''
  Nucl.\ Phys.\  B {\bf 241}, 493 (1984).
  
  
\bibitem{Affleck:1984xz}
  I.~Affleck, M.~Dine and N.~Seiberg,
  ``Dynamical Supersymmetry Breaking In Four-Dimensions And Its
  Phenomenological Implications,''
  Nucl.\ Phys.\  B {\bf 256}, 557 (1985).


\bibitem{MN}
  H.~Murayama and Y.~Nomura,
  ``Gauge mediation simplified,''
  Phys.\ Rev.\ Lett.\  {\bf 98}, 151803 (2007)
  [arXiv:hep-ph/0612186].

\bibitem{Intriligator:2007py}
  K.~A.~Intriligator, N.~Seiberg and D.~Shih,
  ``Supersymmetry Breaking, R-Symmetry Breaking and Metastable Vacua,''
  JHEP {\bf 0707}, 017 (2007)
  [arXiv:hep-th/0703281].

\bibitem{Abel:2007nr}
  S.~A.~Abel, C.~Durnford, J.~Jaeckel and V.~V.~Khoze,
  ``Patterns of Gauge Mediation in Metastable SUSY Breaking,''
  JHEP {\bf 0802}, 074 (2008)
  [arXiv:0712.1812 [hep-ph]].


\bibitem{Raby:2008zz}
  S.~Raby,
  ``Grand Unified Theories,''
{http://www.slac.stanford.edu/spires/find/hep/www?irn=8096635}{SPIRES
entry}

\bibitem{decayreview}
  C.~Amsler {\it et al.}  [Particle Data Group],
  ``Review of particle physics,''
  Phys.\ Lett.\  B {\bf 667} (2008) 1.

\bibitem{decayreview2}B.~Bajc, P.~Fileviez Perez and G.~Senjanovic,
``Minimal supersymmetric SU(5) theory and proton decay: Where do
we stand?,'' arXiv:hep-ph/0210374.

\bibitem{Goto:1998qg}
  T.~Goto and T.~Nihei,
  ``Effect of RRRR dimension five operator on the proton decay in the minimal
  SU(5) SUGRA GUT model,''
  Phys.\ Rev.\  D {\bf 59}, 115009 (1999)
  [arXiv:hep-ph/9808255].

\bibitem{Murayama:2001ur}
  H.~Murayama and A.~Pierce,
  ``Not even decoupling can save minimal supersymmetric SU(5),''
  Phys.\ Rev.\  D {\bf 65}, 055009 (2002)
  [arXiv:hep-ph/0108104].



\end{thebibliography}
\end{document}